\documentclass[epj]{svjour}
\usepackage{epsfig} 
\usepackage{times}

\begin{document}
%
%
\title{Regge approach to charged-pion photoproduction at invariant energies
  above 2 GeV}
\author{
A.~Sibirtsev\inst{1,2}, J.~Haidenbauer\inst{3}, S.~Krewald\inst{3}, 
T.-S.H.~Lee\inst{1,4}, U.-G.~Mei{\ss}ner\inst{2,3} and A.W. Thomas\inst{1,5,6}} 
\institute{Excited Baryon Analysis Center (EBAC), Thomas Jefferson National Accelerator
Facility, Newport News, Virginia 23606, USA \and
Helmholtz-Institut f\"ur Strahlen- und Kernphysik (Theorie), 
Universit\"at Bonn, Nu\ss allee 14-16, D-53115 Bonn, Germany \and
Institut f\"ur Kernphysik (Theorie), Forschungszentrum J\"ulich,
D-52425 J\"ulich, Germany  \and Physics Division, Argonne National Laboratory,
Argonne, Illinois, 60439, USA
\and Theory Center,  Thomas Jefferson National Accelerator
Facility, 12000 Jefferson Ave., Newport News, Virginia 23606, USA
 \and College of William and Mary, Williamsburg, VA 23187, USA}
\date{Received: date / Revised version: date}

\abstract{
A Regge model with absorptive corrections is employed in 
a global analysis of the world data on positive and negative pion 
photoproduction for photon energies from 3 to 8~GeV. In this region
resonance contributions are expected to be negligible so that 
the available experimental information on differential cross sections and 
single polarization observables at $-t{\leq}2$ GeV$^2$ 
allows us to determine the non-resonant part of the reaction amplitude
reliably.
The model amplitude is then used to predict observables for photon energies
below $3$ GeV. Differences between our predictions and data in 
this energy region are systematically examined as possible signals for 
the presence of excited baryons. 
We find that the data available for the polarized 
photon asymmetry show promising 
resonance signatures at invariant energies around 2~GeV. 
With regard to differential 
cross sections the analysis of negative pion photoproduction data,
obtained recently at 
JLab, indicates likewise the presence of resonance structures around 2~GeV. 
}


\authorrunning{ A.~Sibirtsev {\it et al.} }
\titlerunning{Regge approach to charged-pion photoproduction}

\maketitle
\section{Introduction}
The generation of hadron mass including the excited baryon
spectrum~\cite{Isgur,Capstick,Capstick1,Capstick2,Shifman,Ioffe1,Meissner1,Ioffe2}
is one of the unsolved puzzles of QCD that explicitly involves such
fundamental properties as chiral symmetry and confinement. Historically, and on
a more phenomenological level, there are two different approaches to hadron mass 
generation. The first, beginning with the
Gell-Mann-Levy sigma model~\cite{GellMann} and the Nambu-Jona-Lasino
model~\cite{Nambu} has the mass originating from the spontaneous breaking of chiral
symmetry. An alternative approach considers mass generation in terms of the energy
accumulation in the string connecting color charges, which results in the
very successful phenomenology of Regge trajectories for high lying
baryons~\cite{Barger1,Barger2,Collins4,Iwasaki,Iachello,Robson,Kirchbach1,Kirchbach2}.
On a more fundamental level, the generation of mass in QCD is related to the
anomalous breaking of the scale invariance of the classical gauge theory in
terms of the trace anomaly \cite{Collins:1976yq,Ji:1994av}.
This anomaly clearly shows that hadrons
made of light quarks acquire the bulk of their mass from field (binding)
energy.

A rough inspection of the excited baryon spectrum as given by the Particle Data
Group~\cite{Yao} suggests an impressive regularity for nucleon and Delta states
above $\simeq$1.8~GeV. The states with the same spin but opposite parity are
almost degenerate. At the same time, a parity doubling is not observed for the 
well established low lying baryons. Unfortunately, the PDG has only 
assigned many of these 
observed states one or two stars and some of the doublet
partners for the baryons with masses above 2~GeV have not been observed because
the spectroscopy of  high lying baryons is a non-trivial problem. 
Therefore the crucial question of whether the parity doubling of the high mass
baryons has systematic nature remains open. 

Obviously, one can equally well ask
why parity doubling was not
observed for low mass baryons and what is the QCD symmetry behind this
phenomenon? Only recently it was
proposed~\cite{Glozman1,Glozman2,Jido1,Jido2,Cohen1,Cohen2,Cohen3,Jaffe1,Jaffe2}
that parity
doubling might reflect the  restoration of spontaneously broken chiral symmetry of QCD.
A clear testable prediction of chiral symmetry restoration
is the existence of chiral partners of those high-lying states with a 4-stars
rating\footnote{The $G_{17}$ baryon with mass of $\simeq$2.19 GeV and
spin $J{=}7/2$ is quoted~\cite{Yao} with four stars and has negative
parity. The $I_{1,11}$ has mass $\simeq$2.6~GeV, spin $J{=}11/2$,
negative parity and is quoted with three stars.}, namely the $N(2190)$ and
$N(2600)$. The parity partners of those established states are presently 
missing in the known baryon spectrum. Note that there are also missing chiral partners 
of $N$ and $\Delta$ baryons, rated with less than three stars, in the mass region from 
2.2 to 3 GeV as listed, for instance, in Ref.~\cite{Gonzalez}.

However, these speculations about chiral symmetry restoration in the spectrum
are not the only way to explain the apparent doubling phenomenon. It was shown
in the framework of a covariant constituent quark model
\cite{Loring:2001kx,Loring:2001ky}, that the instanton
induced multi-fermion interaction leads to a lowering of selected states that
accidentally become degenerate with their parity partners \cite{Loring:2001bp}.

Despite the considerable amount of $\pi{N}$ data available at invariant collision
energies $\sqrt{s}{\ge}2$~GeV, the high-mass baryon spectrum has never been
systematically explored. The known nucleon and $\Delta$ resonances with masses above
2~GeV were found already in the early single channel $\pi{N}{\to}\pi{N}$ partial wave
analyses~\cite{Hoehler1,Hoehler2,Koch,Cutkovsky,Hendry}. The results of the 1990
analysis~\cite{Manley} of $\pi{N} \to \pi{N}$ and $\pi{N}{\to}\pi\pi{N}$ data
are in reasonable agreement with the previous
findings~\cite{Hoehler1,Hoehler2,Koch,Cutkovsky,Hendry}. The most recent GWU
analysis~\cite{Arndt96,GWU} of $\pi{N}$ scattering covers now 
energies up to $\sqrt{s}{=}2.6$~GeV. However, the description of the data deteriorates
noticeably above $\simeq$2.4 GeV, which is reflected in a sharp increase in the 
achieved $\chi^2$. Unfortunately, the status of high-mass
resonances has not yet been settled. Furthermore, the available $\pi{N}$ data 
base at $\sqrt{s}{\ge}1.8$~GeV is far from being complete. Specifically, for
a conclusive analysis with regard to excited baryons additional polarization 
data are necessary. But it is rather difficult to perform the experiments in
question in the near future because of the lack of suitable pion beams.

Fortunately we can use electromagnetic beams to study the 
excited baryons with masses above 1.8~GeV. The
high-energy beams required (with $E_\gamma \geq 1.3$ GeV in the laboratory
frame) are available at JLab, ELSA, GRAAL, SPring-8 and 
the new MAMI-C project. Data from these facilities on photo- and
electro production of pseudoscalar and vector mesons should allow us 
to extract nucleon resonance parameters associated with excited 
baryon states. Among the various reaction channels with different final 
states, single-pion photo-production provides
the most straightforward access to baryon spectroscopy. 
This reaction is the focus of the present work and we study it 
within a Regge approach. 

As far as theoretical investigations are concerned 
$\pi N$ dynamics in the so-called fourth resonance region, 
i.e. for energies $2 \le \sqrt{s} \le 3$ GeV, is practically uncharted 
territory. 
Most of the existing studies within the conventional 
meson-exchange picture, utilizing phenomenological Lagrangians, 
are restricted to energies up to the $\Delta$ (1232) excitation 
region, cf. ~\cite{leesmith-07} for a recent review. 
There are only very few meson-exchange models that considered
$\pi{N}$ scattering up
to $\sqrt{s}{\simeq}2$~GeV \cite{julich1,julich2,Yang2006,Chen2007,Julia2007}.
Also, with regard to single pion photoproduction, the majority of the
investigations cover only the energy region up to 
$\sqrt{s}{\simeq}1.5 $~GeV \cite{Gross2,Sato,Fuda,Pascalutsa,Haberzettl}. 
A coupled-channel approach 
for the analysis of the data of photo- and electro-production of $\pi N$, 
$\eta N$, and $\pi\pi N$ final states up to $\sqrt{s}{\leq}2$ GeV is 
formulated in Ref.~\cite{msl}. Finally, on a more phenomenological level, 
$K$-matrix based coupled-channel analyses of pion and photon induces 
reactions up to energies $\sqrt{s}{\simeq}2$~GeV were presented in 
Refs.~\cite{Feuster1,Penner,Klempt,Klempt1}. 

A description of the $\pi N$ system within such meson-exchange
models becomes very complex and difficult in practice at higher energies 
$\sqrt{s}{\geq}2$ GeV, say.
Obviously, the number of reaction channels and the number of exchange 
diagrams, which define the basic interactions of these models, 
increases tremendously and most of the pertinent parameters are not known well.
In contrast, the Regge 
model~\cite{Collins4,Perl,Collins2,Collins3,Caneschi,Matthiae,Levin}, 
with a relatively 
transparent parameterization of the reaction amplitude, has been fairly successful in 
describing hadron scattering at high energies, i.e. for photon energies above 3 GeV, 
say. Therefore, naturally the question arises whether the information contained in the 
Regge model can be exploited for investigations at lower energies and, in particular, 
in the transition region to those energies where the analysis of pion photoproduction 
data based on meson-exchange models might be still tractable. 
This issue is the objective of the present work.
 
In the present paper we utilize the Regge formalism 
to perform a global analysis of the world data set on charged-pion
photoproduction in the high photon-energy region where resonance 
contributions are expected to be negligible or absent and thus 
the non-resonant part of the reaction amplitude can be determined 
reliably. In fitting the experimental results it is important
to note that the Regge approach is applicable only in the
small momentum transfer region. Thus, it is natural 
that the Regge model gradually fails to reproduce the data 
as the momentum-transfer increases. In our global analysis we therefore 
include data on differential cross sections and single and double 
polarization observables in the energy range 3${\leq}E_\gamma{\leq}$8 GeV 
but with the restriction $-t{\leq}2$ GeV$^2$. The data considered were 
all obtained around or before 1980. 

Once the parameters of the Regge model are fixed the corresponding amplitudes
are extended to lower energies. Specifically, they are used to 
compute observables in the energy region $1.4 \le E_\gamma \le 3$ GeV,
that corresponds roughly to invariant energies $2 \le \sqrt{s} \le 2.6$ GeV,
and the results are confronted 
with data in this energy region, for example with the differential
cross sections for charged pion photoproduction measured 
recently~\cite{Zhu,Zhu1} in Hall A at JLab. 
In this energy region
differences between our predictions and the data are expected. 
But these differences are precisely what we are after because they 
could be a signal for the presence of resonances and, thus, could be used 
to identify excited baryon states with masses $\sqrt{s}{\geq}2$ GeV.
Consequently, we explore at which $\sqrt{s}$ the presently available data 
possibly show room for additional resonance contributions and we examine the 
issue of which observables are the most crucial ones for excited 
baryon spectroscopy. 

The paper is organized as follows. The formulation of the model is given in
Section~2. An analysis of positive and negative pion photoproduction is given in
Sections 3 and 4, respectively. In Section 5 we consider the $\pi^-/\pi^+$
production ratio. The paper ends with a discussion of further 
perspectives both in experiment and theory.

\section{The Model}

Because the mechanisms of charged and neutral pion photoproduction
are different, we will only analyze 
the data for the $\gamma{p}{\to}\pi^+{n}$ and 
$\gamma{n}{\to}\pi^-{p}$ reactions. Indeed for these reactions pion
exchange dominates at small $-t$ whereas $\omega$-exchange is forbidden
altogether, while the situation is opposite for $\pi^0$-meson
photoproduction. However, as was emphasized in Ref. \cite{Rahnama2} and will
be discussed later, it is already a highly non-trival task to obtain a 
Regge model fit to all of the considered world data of charged pion 
photoproduction.

The previous
phenomenological analyses~\cite{Rahnama2,Storrow1,Kellett} of single charged
pion photoproduction at high photon energies clearly indicate that
a pure Regge pole model can not give an accurate description of the data. 
For example, it is known experimentally that the differential cross section 
increases when the four-momentum transfer approaches $t{=}0$. However, 
while the reaction is certainly dominated by pion exchange for small four-momentum 
transfer $|t|{<}m_\pi^2$, where $m_\pi$ is the pion mass, the pion-exchange 
contribution alone cannot explain the data because it vanishes when $|t|$ 
approaches zero.

To resolve the problem it was proposed to include a ``pion parity
doublet'' \cite{Ball1,Henyey1} which allows a good description of the 
available forward differential cross section data. But the same model
could not reproduce the polarization data. Attempts have been
made~\cite{Rahnama2,Storrow1,Ball,Jackson,Jackson1,Froyland,Rahnama1} to 
include absorptive corrections. Using a poor man's absorption
correction~\cite{Rahnama2,Storrow1} to the pion exchange
enabled a good fit to the data at small $|t|$. However, a further
inspection~\cite{Kellett,Guidal,Vanderhaeghen} indicated that the 
differential cross section of charged pion photoproduction increases 
too sharply to be explained only by the interference between the 
pion-exchange and Regge cut contributions.

A good description of the sharp forward peaks observed in charged pion 
photoproduction, while satisfying also gauge invariance, was achieved 
by a proper inclusion of nucleon ($s$ or $u$ channel) exchange. 
The calculations based on this approach~\cite{Kellett} also reproduced 
the photon asymmetry data. Unfortunately such a gauge-invariant 
unitarized Regge model was not applied to perform a
systematic analysis of the world  data of
charged pion photoproduction reactions. 
In this work, we will make progress in this direction.

To have a simple approach to describe the forward peaks of
charged pion photoproduction differential cross sections, we do not 
reggeize the pion exchange. Instead, we follow previous works by
assuming~\cite{Kellett} that it contributes as a fixed pole via the 
electric Born term. The resulting amplitude~\cite{Dombey1,Dombey2} 
satisfies gauge invariance. This approach also reduces the number of 
parameters to be determined by a fit to the data. Such a 
gauge invariant amplitude for pion exchange has been employed in
Refs.~\cite{Kellett,Guidal,Kellett1,Blackmon,Kramer1,Berends1,Berends2}. 

Before describing our model in detail let us first mention here
some other problems in the previous analyses.
Quite reasonable agreement between the Regge
model calculations~\cite{Rahnama2,Storrow1} and data was obtained by
incorporating the finite-energy sum rules (FESR) into
the fitting. The use of the FESR requires reliable multipole amplitudes
of pion photoproduction in the whole resonance region. 
Existing partial-wave analyses (PWA)~\cite{Hoehler1,Hoehler2,Koch,Manley}
of pion-nucleon scattering have identified
baryon resonances with masses up to
3~GeV. Presently the PDG listing~\cite{Yao} includes four
baryons with a 4-star rating in the mass region from 2 to 2.5~GeV.
Furthermore, the FESR applied to the $\pi^-p{\to}\pi^0n$
reaction distinctly illustrates~\cite{Barger1,Barger2,Dolen} that the resonance
region extends up to $\sqrt{s}$=3~GeV. However, the most recent 
GWU PWA~\cite{Said3} for pion photoproduction is valid only
for $\sqrt{s}$ below 2.6~GeV. Therefore, an incorporation of FESR into
the analysis of pion photoproduction at high energies
$\sqrt{s} \geq 3$ GeV seems to be impossible at present. We thus 
do not include the FESR in our analysis.

Guided by the previous works described above,
we here develop a gauge invariant Regge model, which 
combines the Regge pole and
cut amplitudes for $\rho$, $a_2$ and $b_1$ exchanges as well as a 
gauge invariant pion-exchange Born term. Indeed
at high energies the interactions before and after the basic Regge
pole exchange mechanisms are
essentially elastic or diffractive scattering described by Pomeron exchange.
Such a scenario can
 be related to the distorted wave approximation and provides a
well defined formulation~\cite{Irving,Sopkovitch,Gottfried,Jackson3,Worden}
for constructing Regge cut amplitudes. 
This approach, which
 can also be derived in an eikonal
formalism~\cite{Arnold} with $s$-channel unitarity~\cite{Jackson},
is used in our work. 
Detailed discussions about the non-diffractive multiple scattering corrections
involving intermediate states which differ from the
initial and final states
and the relevant Reggeon unitarity equations are given in
Refs.~\cite{Irving,Gribov,White3,White4}.
For simplicity we do not consider these much more involved
mechanisms which would increase significantly the number of 
parameters to be fitted.

\subsection{General structure}

In our analysis we use the $t$-channel parity conserving helicity amplitudes
$F_i$ ($i= 1,\ldots ,4$). The $F_i$ have proper crossing and analytic
properties and definite spin-parity in the $t$-channel. $F_1$ and $F_2$ are
the natural and unnatural spin-parity $t$-channel amplitudes to all orders 
in $s$, respectively. $F_3$ and $F_4$ are the natural and unnatural 
$t$-channel amplitudes to leading order in $s$. The amplitudes for 
charged pion photoproduction in the standard isospin decomposition are
given by \cite{Chew}:
\begin{eqnarray}
F^{\gamma{p}{\to}\pi^+{n}}_i=\sqrt{2}[F^0_i + F^-_i], \nonumber \\
F^{\gamma{n}{\to}\pi^-{p}}_i=\sqrt{2}[F^0_i - F^-_i] \ .
\label{gpar}
\end{eqnarray}
The correspondence between different Regge exchanges with $J{\le}2$ and
the amplitudes $F^0_i$ and $F^-_i$ that enter into Eq.~(\ref{gpar}) is given
in Table~\ref{traj}. 

\begin{table}
\begin{center}
\caption{Correspondence between $t$-channel Regge exchanges and the helicity 
amplitudes $F_i^-$ and $F_i^0$ ($i{=}1-4$). Here $P$ is parity, $J$ the spin, $I$
the isospin, $G$ the $G$-parity, $\cal N$ the naturalness and $\cal S$ the
signature factor.}
\label{traj}
\begin{tabular}{|c|c|c|c|c|c|c|c|}
\hline\noalign{\smallskip}
      & $P$  & $J$   & $I$  & $G$ & $\cal N$& $\cal S$& Exchange \\
\noalign{\smallskip}\hline\noalign{\smallskip}
$F_1^-$  & +1 & 2 &  1 & $-1$ & +1 & +1 & $a_2$  \\
$F_1^0$ & $-$1 & 1 & 1 & +1 & +1 & $-$1 &  $\rho$  \\
$F_2^-$ & $-1$ &  0 & 1 & $-1$ & $-1$ & +1 & $\pi$ \\
$F_2^0$ & +1 &  1 & 1 & +1 & $-1$ & $-1$ & $b_1$ \\
$F_3^-$ & +1 & 2 &  1 & $-1$ & +1 & +1 & $a_2$  \\
$F_3^0$ & $-1$ & 1 & 1 & +1 & +1 & $-1$ &  $\rho$  \\
$F_4^-$ & +1 & 1 & 1 & $-1$ & $-1$ & $-1$ & $a_1$ \\
$F_4^0$ & $-1$ & 2 & 1 & +1 & $-1$ & +1 & $\rho_2$\\
\hline\noalign{\smallskip}
\end{tabular}
\end{center}
\end{table}

Both natural and unnatural parity particles can be exchanged in the $t$-channels
in charged pion photoproduction provided they have isospin $I{=}1$ and
$G$-parity $G{=}{\pm}1$. The naturalness $\cal N$ for
natural (${\cal N}{=}{+1}$)  and unnatural (${\cal N}{=}{-1}$)  parity exchanges
is defined as
\begin{eqnarray}
{\cal N}=+1 \,\,\, \mathrm{if} \,\,\, P=(-1)^J, \nonumber \\
{\cal N}=-1 \,\,\, \mathrm{if} \,\,\, P=(-1)^{J+1},
\end{eqnarray}
where $P$ and $J$ are the parity and spin of the particle, respectively.
Furthermore in Regge theory each exchange is denoted by a signature
factor ${\cal S}{=}{\pm}1$ defined as~\cite{Irving,Collins2,Collins3}
\begin{eqnarray}
{\cal S} = P \times {\cal N} = (-1)^J.
\label{signat}
\end{eqnarray}

\subsection{Observables}
\label{obs}
The relation between the $t$-channel helicity amplitudes $F_i$ 
and the observables can be constructed via the transformation to the
$s$-channel helicity amplitudes $S_1$, $S_2$, $N$ and $D$.
Following Wiik's abbreviations~\cite{Wiik},  $S_1$ and $S_2$ are single spin 
flip amplitudes, $N$ is the spin non-flip and $D$ is the double spin flip 
amplitude, respectively.
The asymptotic crossing relation, which is useful for the analytical
evaluation of the helicity amplitudes, is given by
\begin{eqnarray}
\!\!\!\left[
\begin{matrix}{F_1 \crcr F_2 \crcr F_3 \crcr F_4}
\end{matrix}
\!\right]\!\!{=}\frac{-4\sqrt{\pi}}{\sqrt{-t}}\!\!
\left[\begin{matrix}{\!2m & \sqrt{-t} & -\sqrt{-t} & \!\!2m \crcr 
0 & \sqrt{-t} & \sqrt{-t} & 0 \crcr 
t & \!\!2m\sqrt{-t} & \!\!{-}2m\sqrt{-t} & t \crcr 
1 & 0 & 0 & {-}1 }
\end{matrix}\!\!
\right]\!\!
\left[
\begin{matrix}{S_1 \crcr N \crcr D \crcr S_2}
\end{matrix}
\right] .
\label{matr1}
\end{eqnarray}

Utilizing the relations of Ref.~\cite{Baker} the 
$\gamma{p} \to \pi^+{n}$ and $\gamma{n} \to \pi^-{p}$ 
observables analyzed in the present study are given by 
\begin{eqnarray}
\frac{d\sigma}{dt}&=&\frac{1}{32\pi}\left[ \frac{
t|F_1|^2-|F_3|^2}{(t-4m^2)} +|F_2|^2-t|F_4|^2\right]~, 
\label{obs1} \\
\frac{d\sigma}{dt}\Sigma&=&\frac{1}{16\pi}\left[ \frac{
t|F_1|^2-|F_3|^2}{(t-4m^2)} -|F_2|^2+t|F_4|^2\right] ~,
\label{obs2} \\
\frac{d\sigma}{dt}T&=&\frac{\sqrt{-t}}{16\pi}\,\, {\rm Im} \left[ \frac{
-F_1 F_3^\ast}{(t-4m^2)} + F_4F_2^\ast \right] ~,
\label{obs3} \\
\frac{d\sigma}{dt}R&=&\frac{\sqrt{-t}}{16\pi}\,\, {\rm Im} \left[ \frac{
-F_1 F_3^\ast}{(t-4m^2)} - F_4F_2^\ast \right] \ , 
\label{obs4} 
\end{eqnarray}
where the appropriate isospin combinations of the $F_i$'s according 
to Eq.~(\ref{gpar}) need to be taken. The relations in 
Eqs.~(\ref{obs1}) - (\ref{obs4}) allow one 
to obtain constraints for the $t$-channel
helicity amplitude directly from experimental observables. Note that
Eq.~(\ref{matr1}) is appropriate only at $s{\gg}t$, since it does not 
account for the higher order corrections that are proportional to $t/4m^2$. 
The amplitudes $F_i$ are related to the usual CGLN 
invariant amplitudes $A_i$ \cite{Chew} by 
\begin{eqnarray}
F_1&=&-A_1+2mA_4~, \nonumber \\
F_2&=&A_1+tA_2~, \nonumber \\
F_3&=&2mA_1-tA_4~, \nonumber \\
F_4&=&A_3 \ . 
\label{invari}
\end{eqnarray}
Expressions for the experimental observables in terms of the
amplitudes $A_i$ are listed, for instance, in Ref.~\cite{Berends}. 
The often used multipole amplitudes can be constructed from
the helicity amplitudes using the relations given in Ref.~\cite{Said1}. 
		
\subsection{Structure of the amplitudes}

The pion photoproduction amplitude of our model is given by 
\begin{eqnarray}
F_i=F^{(\pi)}_i+F^{(Regge)}_i \ , 
\label{eq:ampt}
\end{eqnarray}
where contributions from Regge exchanges of the $\rho$, $a_2$ and $b_1$
trajectories 
are taken into account. Their concrete structure and parameterization is 
described in detail in the next subsection. As mentioned, the contribution 
from pion exchange is treated differently and will be discussed and 
described in detail in a separate subsection below. 

\subsubsection{Regge amplitudes}

Similar to the particle-exchange Feynman diagram, each reaction 
amplitude $F$ is factorized in terms of a propagator $G$ and a vertex
function $\beta$
\begin{eqnarray}
F_i^{(Regge)}(s,t) \sim \beta_i \times  G \,.
\label{eq:f}
\end{eqnarray} 
However, there is a difference in defining the propagator. 
The basic reaction mechanism in the Regge model is not associated with 
the exchange of certain particles but with the exchange of certain quantum 
numbers. Therefore, the mass of the exchanged particle does not appear 
in the amplitudes explicitly. Accordingly, 
the usual Feynman propagator, which contains the mass $m$ of the
exchange particle, is replaced by the Regge propagator 
\begin{eqnarray}
G\sim \frac{1}{t-m^2} \Rightarrow  \frac{1 {+} {\cal S}\exp[-i \pi \alpha
(t)]}{\sin[\pi\alpha(t)] \,\,\Gamma[\alpha(t){+}1]}
\left[\frac{s}{s_0}\right]^{\alpha(t)-1},
\label{rpropa}
\end{eqnarray}
where $s_0$=1~GeV$^2$ is a parameter for defining a dimensionless
amplitude, ${\cal S}$ is the signature factor given in Table~\ref{traj} 
and $\alpha(t)$ is the Regge trajectory. The trajectories are
the most essential part of the Regge model and they are defined 
by the spins ($J$) and masses ($m_J$) of the particles 
with a fixed G-parity, ${\cal N}$ and ${\cal S}$. Specifically, the function 
$\alpha(t)$ characterizing the trajectory is obtained from the relation 
$\alpha(m_J)$ = $J$ applied to those particles that form the trajectory. 
The trajectories pertinent to our approach will be discussed below.

Obviously the Regge propagator of 
Eq.~(\ref{rpropa}) accounts for the whole family of particles or poles, which lie 
on a certain trajectory, where the trajectory is {\it named after} the lowest 
$J$ state. Thus, by 
considering different trajectories constructed in the unphysical $t{\ge}$0 region
one can effectively include all possible exchanges allowed by the conservation of
quantum numbers. This is an obvious advantage of the Regge theory, since with
increasing energy it is necessary to include the exchanges of higher-mass and
higher-spin particles and a description within standard relativistic meson-exchange 
models would become too involved or even unmanageable.

From Eq.~(\ref{rpropa}), we see that the factor $\sin[\pi\alpha(t)]$ would generate 
also poles at $t{\le}0$ when $\alpha(t)$ assumes the values $0,-1,\ldots$. 
The function $\Gamma[\alpha(t){+}1]$ is {\it introduced} to suppress those 
poles that lie in the scattering region because 
\begin{eqnarray}
\frac{1}{\Gamma[\alpha(t)+1]} = -\frac{\sin [\pi \alpha(t)]}{\pi}
\,\Gamma(\alpha(t)).
\label{eq:supp}
\end{eqnarray}
However, the suppression of the poles in the physical region can be done by 
other means too. This issue will be discussed below when we introduce the
concrete parameterization of the Regge amplitudes that we use. 

The structure of the vertex function $\beta$ of Eq.~(\ref{eq:f})
is defined by the quantum numbers of the particles at the interaction
vertex, similar to the usual particle exchange Feyman diagram. 
This vertex function is taken to be real and hence
$\rho$, the ratio of the real to imaginary parts of the reaction
amplitude, is given by 
\begin{eqnarray}
\rho = \frac{{\rm Re}~ F}{{\rm Im}~  F} \propto -\frac{{\cal S}+\cos[\pi \alpha(t)]}
{\sin[\pi \alpha(t)]},
\end{eqnarray}
for a specific Regge exchange, i.e. simply by the phase of Eq. (\ref{rpropa}). 
This phase is required by the fixed-$t$ dispersion relation and is well verified 
experimentally\footnote{For instance, the ratio $\rho$ can be measured directly
in forward elastic scattering.}.

The Regge amplitudes used in our model calculation are of the
form
\begin{eqnarray}
F^{(Regge)}_i(s,t){=}\!\!\sum_{j}[ F^{(pole)}_{ij}(s,t){+}F^{(cut)}_{ij}(s,t)], 
\label{eq:regge-p}
\end{eqnarray}
where $j=1,2,3$ denote the trajectories $a_2$, $\rho$ and $b_1$,
respectively. (Note that we do not consider the amplitude $F_4$ and the
corresponding trajectories $a_1$ and $\rho_2$ in the present investigation 
for reasons that are discussed later.)
Each of the pole amplitudes are
parameterized as~\cite{Irving} (suppressing the subscripts $ij$)
\begin{eqnarray}
F^{(pole)}(s,t)=\beta (t) \frac{1 {+} {\cal S}\exp[-i \pi \alpha (t)]}{\sin[\pi
\alpha(t)]}
\left[\frac{s}{s_0}\right]^{\alpha(t)-1},
\label{amplibas}
\end{eqnarray}
where $\beta(t)$ is the residue function which
accounts for the $t$-dependence and
the coupling constant at the interaction vertex, and
${\cal S}$ is the signature factor given by Eq.~(\ref{signat}) 
and listed in Table~\ref{traj}.

The residue functions $\beta(t)$  used in our analysis
are compiled in Table~\ref{tab0}. 
They are similar to the ones used in some of the previous
analyses~\cite{Rahnama2,Kellett}. The factor
$\alpha(t)[\alpha(t){+}1]$ in Table ~\ref{tab0}
is used to suppress the poles of the
propagator in the scattering region. Alternatively
this suppression can be achieved~\cite{Irving} by
introducing the $\Gamma[\alpha(t)]$ function as seen in
Eqs.~(\ref{rpropa}) and (\ref{eq:supp}).
One can also introduce a factor
$[\alpha(t){+}n]$ with $n{=}2,3,\dots$  to suppress poles at large $-t$.
However, we do not apply the Regge model beyond $|t|{=}$2~GeV, and therefore 
such a suppression factor is not considered. 
We should mention that in some studies~\cite{Chiu} 
it was proposed to drop the $\alpha(t)$ factor for the $\rho$ pole exchange. 
But we keep this factor in our model. In fact, we found that it has 
practically no influence on the achieved $\chi^2$ of the fit.

\begin{table}
\begin{center}
\caption{Parameterization of the $\beta(t)$ functions for the amplitudes
$F_i$, \,  ($i{=}1{-}3$). Here $c_ {ij}$ is the coupling constant 
where the double index refers to the amplitude and the type of exchange, 
as specified in the Table, while $\alpha_j(t)$ denotes the trajectory for the 
type of exchange. These trajectories are given by 
Eqs.~(\ref{traj1}) and (\ref{traj2}). 
}
\label{tab0}
\begin{tabular}{|l|l|l|l|}
\hline\noalign{\smallskip}
\multicolumn{4}{|c|}{ Pole amplitudes }\\
\hline\noalign{\smallskip}
      & Residue function $\beta(t)$ & Exchange & $j$ \\
\noalign{\smallskip}\hline\noalign{\smallskip}
$F_1$ & $c_{11}\, \alpha_1(t)\, [\alpha_1(t){+}1]$ & 
$a_2$ & 1 \\ 
$F_1$ & $c_{12} \,\alpha_2(t)\, [\alpha_2(t){+}1]$ & $\rho$ & 2 \\ 
\noalign{\smallskip}\hline\noalign{\smallskip}
$F_2$ & $c_{23}\, t \,\alpha_3(t)\, [\alpha_3(t){+}1]$  &
$b_1$ & 3 \\ 
\noalign{\smallskip}\hline\noalign{\smallskip}
$F_3$ & $c_{31}\, t \, \alpha_1(t)\, [\alpha_1(t){+}1]$  &
$a_2$ & 1 \\ 
$F_3$ & $c_{32} \,t \, \alpha_2(t)\,[\alpha_2(t){+}1]$  & $\rho$
& 2 \\ 
\noalign{\smallskip}\hline\noalign{\smallskip}
\multicolumn{4}{|c|}{ Cut  amplitudes }\\
\hline\noalign{\smallskip}
$F_1$ & $c_{14}\, [\alpha_4(t){+}1] \, \exp[d_4t]$ & 
$a_2$ & 4\\ 
$F_1$ & $c_{15} \, \alpha_5(t)\, \exp[d_5t]$ &
$\rho$ & 5 \\ 
$F_1$ & $c_{16}\, [\alpha_6(t){+}1] \, \exp[d_6t]$  &
$b_1$ & 6 \\ 
\noalign{\smallskip}\hline\noalign{\smallskip}
$F_2$ & $c_{24}\, [\alpha_4(t){+}1]\, \exp[d_4t]$ & 
$a_2$ & 4\\ 
$F_2$ & $c_{25} \, \alpha_5(t)\, \exp[d_5t]$ &
$\rho$ & 5 \\ 
$F_2$ & $c_{26}\, [\alpha_7(t){+}1]\, \exp[d_6t]$  &
$b_1$ & 6 \\ 
\noalign{\smallskip}\hline\noalign{\smallskip}
$F_3$ & $c_{34}\, [\alpha_4(t){+}1]\, \exp[d_4t]$ & 
$a_2$ & 4\\ 
$F_3$ & $c_{35} \, \alpha_5(t)\,   \exp[d_5t]$ &
$\rho$ & 5 \\ 
$F_3$ & $c_{36}\, [\alpha_6(t){+}1]\,  \exp[d_6t]$  &
$b_1$ & 6 \\ 
\noalign{\smallskip}\hline\noalign{\smallskip}
\end{tabular}
\end{center}
\end{table}

The trajectories are of the following linear form:
\begin{eqnarray}
\alpha(t){=}\alpha_0+\alpha^\prime{t} \ , 
\label{eq:traj}
\end{eqnarray}
where the parameters ($\alpha_0$, $\alpha^\prime$)
for the considered $a_2$, $\rho$ and $b_1$ trajectories are taken 
over from analyses of other reactions~\cite{Irving,Sibirtsev2}.
Explicitly we have for the considered $a_2$, $\rho$ and $b_1$ trajectories 
\begin{eqnarray}
\alpha_{a_2} &=& \alpha_1 =0.4{\phantom{2}}+ 0.99 \, t \nonumber \\
\alpha_{\rho_{\phantom{2}}} &=&\alpha_2 = 0.53 + 0.8 \, t \nonumber \\
\alpha_{b_1} &=& \alpha_3 = 0.51 + 0.8\, t \ .
\label{traj1}
\end{eqnarray}

In defining the Regge cut amplitudes $F^{(cut)}$ of Eq.~(\ref{eq:regge-p})
we use the following parameterization based on the
absorption model~\cite{Collins2,White1,White2,Henyey,Kellett}
(suppressing again the subscripts)
\begin{eqnarray}
F^{(cut)}(s,t){=} \frac{\beta (t)}{\log{(s/s_0)}}
 \frac{1 {+}{\cal S}\exp[{-}i\pi \alpha_c (t)]}{\sin[\pi
\alpha_c(t)]}\!\!
\left[\frac{s}{s_0}\right]^{\alpha_c(t)-1}\!\!\!\!\!\!,
\label{eq:trajcut}
\end{eqnarray}
with the trajectories defined by
\begin{eqnarray}
\alpha_c =\alpha_0 +\frac{\alpha^\prime\alpha_P^\prime \, t}
{\alpha^\prime + \alpha_P^\prime}\,,
\label{traj2}
\end{eqnarray}
where $\alpha_0$ and $\alpha^\prime$ were taken from the pole
trajectory given by Eqs.~(\ref{eq:traj}) and (\ref{traj1}),
and $\alpha_P^\prime{=}0.2$~GeV$^{-2}$ is
the slope of the pomeron trajectory.
The residue functions $\beta(t)$ of Eq.~(\ref{eq:trajcut})
 are also listed in
Table~\ref{tab0}, where the relevant cut trajectories are numerated
as $\alpha_4$, $\alpha_5$, $\alpha_6$  for the  $a_2$, $\rho$ and
$b_1$ cut amplitudes, respectively. 

\subsubsection{Pion-exchange amplitude}
\label{pionsec}

As already mentioned in the beginning of this section, 
we do not reggeize the pion exchange. Instead, we follow previous 
works by assuming~\cite{Kellett} that it contributes as a
fixed pole via the electric Born term. Indeed, pion exchange 
dominates the region $-t < m_\pi^2$ and in this region the Regge
propagator can be savely replaced by the Feyman propagator. 
The resulting amplitude~\cite{Dombey1,Dombey2} satisfies gauge invariance 
in photoproduction.
This approach allows us to describe the forward peaks of charged pion 
photoproduction differential cross sections in a simple way. 
It also reduces the number of parameters to be fitted by the data. 

The gauge invariant pion Born term $F^{(\pi)}_i(s,t)$ 
is calculated~\cite{Kellett,Kellett1,Blackmon,Kramer1,Chew,Kramer2} from
the usual pion and nucleon exchange Feynman diagrams for 
$\gamma N \rightarrow \pi N$, but keeping only the
pure electric coupling in the $\gamma{NN}$ vertex. Explicitly,
the invariant amplitudes for the $\gamma{p}{\to}\pi^+n$ reaction are
given by 
\begin{eqnarray}
A_1 &=&- \frac{eg}{s-m_N^2}f(t)~, \nonumber \\
A_2 &=&  \frac{2eg}{(s-m_N^2)(t-m_\pi^2)}f(t)~ \ . 
\label{pionamp}
\end{eqnarray}
(The relation between the invariant and the helicity amplitudes is 
given by Eq.~(\ref{invari}).)
Here $m_N$ and $m_\pi$ stand for the nucleon and pion mass, respectively,
and $e$ and $g$ are the electric and $\pi{NN}$ coupling constants taken as
$e^2/4\pi$=1/137 and $g^2/4\pi$=13.76. 
Following the standard procedure \cite{Kellett} a phenomenologicl form 
factor $f(t)$ is included, 
\begin{eqnarray}
f(t)=a\exp(bt),
\label{fofa}
\end{eqnarray}
where $a$ and $b$ are free parameters to be determined by a fit to the data.
For the $\gamma{n}{\to}\pi^-p$ reaction the gauge invariance of the pion
exchange can be restored by the $u$-channel nucleon exchange~\cite{Dombey} with
invariant amplitudes taken as
\begin{eqnarray}
A_1&=&  \frac{eg}{u-m_N^2}f(t)~, \nonumber \\
A_2&=&- \frac{2eg}{(u-m_N^2)(t-m_\pi^2)}f(t)~.
\label{pionamp1}
\end{eqnarray} 
Indeed for small $-t{\le}m_\pi^2$ where the pion exchange dominates, the
propagators for $s$ and $u$ channels fulfill approximately 
\begin{eqnarray}
-u +m_N^2 \simeq s-m_N^2,
\end{eqnarray}
when neglecting terms smaller than the pion mass squared. That is why in
Refs.~\cite{Kellett,Kellett1} the $u$-channel correction to the gauge
invariance was not specified explicitly and only $s$ channel invariant
amplitudes were given.

As is obvious from Eqs.~(\ref{pionamp}) and (\ref{pionamp1}) in conjunction with 
Eq. (\ref{invari}), the pion-exchange contribution derived above 
con\-trib\-utes to the helicity amplitudes $F_1$ to $F_3$ while the reggeized
pion exchange would contribute only to $F_2$, cf. Table~\ref{traj}. 

\begin{table}[t]
\begin{center}
\caption{Parameters of the model. Here $c_ {ij}$ is the coupling constant
for the $i$th amplitude and the type of exchange, $d_j$ is a cut-off parameter 
for the Regge cut amplitude, while $a$ and $b$ are the parameters of the Born 
term form factor, cf. Table \ref{tab0} and Eq.~(\ref{fofa}).}
\label{tabp}
\begin{tabular}{|l|c|c|c|c|}
\hline\noalign{\smallskip}
$j$ & \multicolumn{3}{|c|}{$c_{ij}$} & $d_j$\\
 & $i{=}1$ & $i{=}2$ & $i{=}3$ & \\
\noalign{\smallskip}\hline\noalign{\smallskip}
1 &$-30.1$ & - & $103.8$ & - \\
2 & $36.1$ & - &$31.0$ & - \\
3 &- &-8.4 &- & - \\
4 & $164.0$ & $-42.0$ & $348.6$ & $1.46$ \\
5 &$-286.8$ & $127.1$ & $-22.1$ & $0.75$ \\
6 &$ 271.9$ & $-141.3$ & $5.9$ & $0.78$ \\
\noalign{\smallskip}\hline\noalign{\smallskip}
 &\multicolumn{2}{c}{$a=$0.8} &\multicolumn{2}{c|}{$b=$1.56} \\
\noalign{\smallskip}\hline\noalign{\smallskip}
\end{tabular}
\end{center}
\end{table}

\subsection{Parameters of the model}
With the formulation presented above the considered reaction amplitude 
has 19 free parameters.
As discussed in section 1, we fix these parameters by a fit to the 
($\pi^+$ and $\pi^-$) production data at energies $3{\leq}E_\gamma{\leq}8 $ 
GeV and momentum-transfers $-t{\leq}2$~GeV$^2$. Some general information on 
those data \cite{Durham} is given in the following sections where the results
are discused.

In order to avoid any dependence of the fit on the starting parameters we used
the random walk method to construct the initial set of parameters and we 
repeated the minimization procedure many times. 
Furthermore, an additional examination was done by
exploring the results for the parameter correlation-matrix 
in order to inspect the stability of the found minimum.

The data for both the positive and negative pion
photoproduction are fitted simultaneously.
The resulting parameters of the model are given in Table~\ref{tabp}. 
The achieved $\chi^2{/}ndf$ amounts to 1.4. We find that some of 
the data from different experiments are slightly inconsistent.
There is no way to improve the confidence level of our global analysis
unless these inconsistent data are removed from the data base.
However, it is difficult to find meaningful criteria for pruning the
data base. 

\subsection{Experimental constraints on ${\bf F_4}$}
\label{f4sec}

As seen in Table \ref{tab0}, we do not include the amplitude $F_4$ in
our analysis. 
The $F_4$ amplitude is given by the $J^{PC}{=}1^{++}$ and $J^{PC}{=}2^{--}$
exchanges and their cuts.  
Those contributions correspond to the exchanges of the $a_1$, $f_1$, $\rho_2$ and 
$\omega_2$ mesons with the indicated quantum numbers\footnote{Formerly the $a_1$
and $\rho_2$ mesons were called $A_1$ and $Z$, respectively~\cite{Irving}},
where the latter two are not well established experimentally~\cite{Yao}.
In addition, the relevant amplitudes are small because the
corresponding trajectories are low-lying in the $J$-plane~\cite{Rahnama3}.
That were the reasons why the $F_4$
amplitude was neglected in many previous studies of charged-pion photoproduction.

By inspecting the relation between the amplitudes $F_i$ and the observables one can
immediately conclude that the single polarization data on target ($T$) and recoil
($R$) asymmetries\footnote{Note that in some publications of experimental 
results the notation is different. Specifically, for the recoil asymmetry 
$P$ is used (instead of $R$) and for the polarized photon asymmetry $A$
(instead of $\Sigma$).} 
are very crucial to determine the role of $F_4$.
Indeed if $T{=}R$ exactly then it follows that $F_4{=}0$. A direct measurement
of the $R{-}T$ difference allows access to $F_4$ in a model independent 
way because
\begin{eqnarray}
R-T = 4\pi \,\sqrt{-t} \,\, {\rm Im} [\,{F_4F^\ast_2}\,]~,
\label{bound1}
\end{eqnarray}
and because the $F_2$ amplitude is well established from the investigation 
of various reactions~\cite{Baker} dominated by $\pi$-meson exchange.
Furthermore, without any model assumption, the following Worden
inequalities~\cite{Worden} can be derived from the relation between the
$F_i$ amplitudes and the observables:
\begin{eqnarray}
|R - T| &\le& 1 - \Sigma~, \label{bound2a}\\
|R + T| &\le& 1 + \Sigma~, \label{bound2b}\\
|D| &\le& \sqrt{1-\Sigma^2}~,
\label{boundc}
\end{eqnarray}
where $D$ denotes the double polarization parameters $G$, $H$, $E$ and $F$,
which are given in Refs.~\cite{Baker,Baker1}. In addition, the observables 
obey the following equations~\cite{Baker,Worden}
\begin{eqnarray}
E^2 + F^2 + G^2 + H^2 &=&1 + R^2-\Sigma^2 - T^2~, \\
FG - EH &=& R - T\Sigma~,
\end{eqnarray}
which allow to construct $F_4$ from the full set of single and double
polarization measurements.
\begin{table}[t]
\begin{center}
\caption{The $\gamma{p}{\to}\pi^+n$ data on differential cross section
analyzed in the present paper. }
\label{tab:data-1}
\begin{tabular}{|r|r|r|r|c|}
\hline\noalign{\smallskip}
\multicolumn{5}{|c|}{ Differential cross section,  $d\sigma{/}dt$ } \\
\hline\noalign{\smallskip}
$E_\gamma$ \,\, & $\sqrt{s}$\,\, & $-t_{min}$  & $-t_{max}$  & Reference  \\ 
(GeV) & (GeV) & (GeV$^2$) & (GeV$^2$) &  \\ 
\noalign{\smallskip}\hline\noalign{\smallskip}
1.1 & 1.72 & 0.47 & 0.71 & \cite{Zhu} \\
1.1 & 1.72 & 4.1$\times$10$^{-3}$  & 0.41 & \cite{Ecklund} \\
1.48 & 1.91 &0.024 & 0.24 & \cite{Buschhorn1,Buschhorn2} \\ 
1.62 & 1.98 & 1.1$\times$10$^{-3}$ & 0.34 & \cite{Buschhorn1,Buschhorn2} \\ 
1.65 & 1.99 & 0.42 & 1.18 & \cite{Zhu} \\
1.77 & 2.05 & 1.3$\times$10$^{-3}$ & 0.38 & \cite{Buschhorn1,Buschhorn2} \\ 
1.8 & 1.99 & 0.47 & 1.32 & \cite{Zhu} \\
1.99 & 2.15 & 1.5$\times$10$^{-3}$ & 0.44 & \cite{Buschhorn1,Buschhorn2} \\ 
2.18 & 2.23 & 1.6$\times$10$^{-3}$ & 0.6 & \cite{Buschhorn1,Buschhorn2} \\ 
2.38 & 2.31 & 1.8$\times$10$^{-3}$ & 0.66 & \cite{Buschhorn1,Buschhorn2} \\ 
2.48 & 2.35 & 0.69 & 1.95 & \cite{Zhu} \\
2.51 & 2.36 & 0.126 & 0.323 & \cite{Dowd} \\ 
2.63 & 2.41 & 2$\times$10$^{-3}$ & 0.74 & \cite{Buschhorn1,Buschhorn2} \\ 
3.25 & 2.64 & 0.124 & 0.45 & \cite{Dowd} \\ 
3.32 & 2.67 & 0.96 & 3.64 & \cite{Zhu} \\
3.4 & 2.69 & 3$\times$10$^{-3}$ & 0.4 &  \cite{Heide}  \\ 
3.4 & 2.69 & 0.09 & 0.33 &  \cite{Joseph} \\ 
3.41 & 2.7 & 0.374 & 1.396 &  \cite{BarYam}\\ 
3.4-4.0 & 2.69-2.9 & 0.25 & 1.31 &  \cite{Elings}\\
4.0 & 2.9 & 0.95 & 5.14 & \cite{Anderson1} \\
4.15 & 2.94 & 2.28 & 4.1 & \cite{Zhu} \\
4.17 & 2.95 & 0.09 & 0.551 & \cite{Dowd} \\ 
4.4 & 3.02 & 6.77 & 7.08 & \cite{Anderson2} \\
5.0 & 3.2 & 0.01 & 0.61 &  \cite{Heide}  \\ 
5.0 & 3.2 & 0.19 & 1.45 &  \cite{Joseph} \\ 
5.0 & 3.2 & 1.44 & 6.8 & \cite{Anderson1} \\
5.0 & 3.2 & 2$\times$10$^{-4}$ & 1.15 & \cite{Boyarski1} \\
5.53 & 3.36 & 4.75 & 5.6 & \cite{Zhu} \\
7.5 & 3.87 & 1.95 & 11.63 & \cite{Anderson1} \\
8.0 & 3.99 & 5.1$\times$10$^{-4}$ & 2.13 & \cite{Boyarski1} \\
11.0 &4.64 & 8$\times$10$^{-4}$ & 2.06 & \cite{Boyarski1} \\
16.0 & 5.56 & 1.2$\times$10$^{-3}$  & 1.95 & \cite{Boyarski1} \\
\noalign{\smallskip}\hline\noalign{\smallskip}
\end{tabular}
\end{center}
\end{table}

Unfortunately, there are no data for both $T$ and $R$ asymmetries available
for the charged-pion photoproduction at photon energies explored in 
the present study. The data~\cite{Geweniger,BarYam2,Sherden} at
$3{\le}E_\gamma{\le}16$~GeV indicate for the photon
polarized asymmetry $\Sigma{\simeq}$0.8. Following Eq.~(\ref{bound2a}) this
implies that $|R{-}T|{\le}0.2$, suggesting that the $F_4$ amplitude could be 
not negligible. The data available for $T$ and $R$ at $E_\gamma{\le}2.25$~GeV
imply that, within the experimental uncertainties, $T{\simeq}R$. At higher 
energies, $E_\gamma{\ge}$2 GeV, experimental results for $T$ and $R$ 
\cite{Bussey1,Bienlein,Booth,Deutsch} are available only for neutral pion
photoproduction. Those suggest also that $T{\simeq}R$. However, the statistical
uncertainty of these data is large and at $E_\gamma{\le}4$~GeV the
comparison of target and recoil asymmetries requires an interpolation in $t$ and
an extrapolation in the photon energy. 

Certainly, apart from the mentioned experimental indications, there are no 
fundamental reasons to ignore the $F_4$ contribution. Indeed, 
we did attempt to include the $F_4$ amplitude in the global fit following the
trajectory parameters given in Ref.~\cite{Irving}. However, it turned out that 
the fit is insensitive to that contribution. In addition, the most crucial data
available~\cite{Genzel,Morehouse} at $E_\gamma{>}$3~GeV for target asymmetries 
are afflicted with large errors. Thus, finally we decided to neglect $F_4$ in 
the present study. 

\section{Results for {\boldmath$\gamma p \rightarrow \pi^+n$}}
\label{secstra}

\begin{table}[t]
\begin{center}
\caption{The $\gamma{p}{\to}\pi^+n$ data on the polarized photon asymmetry
$\Sigma$
(denoted formerly as $A$~\cite{Worden}), target asymmetry $T$, and the recoil 
symmetry $R$ (denoted formerly as $P$~\cite{Worden}) analyzed in the present
paper.}
\label{tab:data-2}
\begin{tabular}{|r|r|r|r|c|}
\hline\noalign{\smallskip}
\multicolumn{5}{|c|}{ Polarized photon asymmetry $\Sigma$ } \\
\hline\noalign{\smallskip}
$E_\gamma$ \,\, & $\sqrt{s}$\,\, & $-t_{min}$  & $-t_{max}$  & Reference  \\ 
(GeV) & (GeV) & (GeV$^2$) & (GeV$^2$) &  \\ 
\noalign{\smallskip}\hline\noalign{\smallskip}
1.55 & 1.95 & 0.15 & 1.39 & \cite{Bussey2} \\
1.65 & 1.99 & 0.16 & 1.5 & \cite{Bussey2} \\
1.95 & 2.13 & 0.2 & 2.08 & \cite{Bussey2} \\
2.25 & 2.26 & 0.23 & 1.89 & \cite{Bussey2} \\
2.5 & 2.36 &0.02 & 0.31 & \cite{Geweniger} \\ 
3.0 & 2.55 &0.15 & 1.16 & \cite{BarYam2} \\ 
3.4 & 2.69 &0.01 & 0.6 & \cite{Geweniger} \\ 
3.4 & 2.69 &2.6$\times$10$^{-3}$  & 0.01 & \cite{Burfeindt} \\ 
5.0 & 3.2 &0.1 & 0.4 & \cite{Geweniger} \\ 
16.0 & 5.56 &5.5$\times$10$^{-3}$  & 1.5 & \cite{Sherden} \\ 
\hline\noalign{\smallskip}
\multicolumn{5}{|c|}{ Target asymmetry $T$ } \\
\hline\noalign{\smallskip}
1.55 & 1.95 & 0.15 & 1.39 & \cite{Bussey2} \\
1.65 & 1.99 & 0.16 & 1.5 & \cite{Bussey2} \\
1.95 & 2.13 & 0.2 & 2.08 & \cite{Bussey2} \\
2.25 & 2.26 & 0.23 & 1.89 & \cite{Bussey2} \\
2.5 & 2.36 &0.1 & 0.87 & \cite{Genzel} \\ 
3.4 & 2.69 &0.1 & 1.14 & \cite{Genzel} \\ 
5.0 & 3.2 &0.1 & 1.25 & \cite{Genzel} \\ 
5.0 & 3.2 &0.019 & 1.02 & \cite{Morehouse} \\ 
16.0 & 5.56 &0.019 & 0.62 & \cite{Morehouse} \\ 
\hline\noalign{\smallskip}
\multicolumn{5}{|c|}{ Recoil asymmetry $R$ } \\
\hline\noalign{\smallskip}
1.55 & 1.95 & 0.15 & 1.39 & \cite{Bussey2} \\
1.65 & 1.99 & 0.16 & 1.5 & \cite{Bussey2} \\
1.95 & 2.13 & 0.2 & 2.08 & \cite{Bussey2} \\
2.25 & 2.26 & 0.23 & 1.89 & \cite{Bussey2} \\
\noalign{\smallskip}\hline\noalign{\smallskip}
\end{tabular}
\end{center}
\end{table}

This section is organized as follows. First we compare the
results based on our model with the data included in our global 
fit, i.e. data in the region 3${\leq}E_\gamma{\leq}$8 GeV and 
$-t{\leq}2$ GeV$^2$. Some general information on those data 
is listed in Tables~\ref{tab:data-1} and \ref{tab:data-2}.
We also confront our model with the available data at the higher energies 
$E_\gamma$=11 and 16 GeV which were not included in our fit.
Information on those data are listed too in the Tables. 

Then we look at data for $1.4{<}E_\gamma{<}$3~GeV where our Regge-model 
results can be considered as predictions. 
The lowest photon energy is chosen in order to cover invariant masses
down to $\sqrt{s}{\simeq}2$~GeV, which is roughly 
the lower end of the fourth resonance region. Note that at these 
energies we definitely expect to be in disagreement with the data. But we
regard such a disagreement as the starting point for exploring possible 
contributions from nucleon resonances. Thus our interest in that 
energy range is to examine systematically for which observables and in 
which kinematical regions discrepancies between our predictions and 
available data occur.

\begin{figure}[t]
\vspace*{-6mm}
\centerline{\hspace*{5mm}\psfig{file=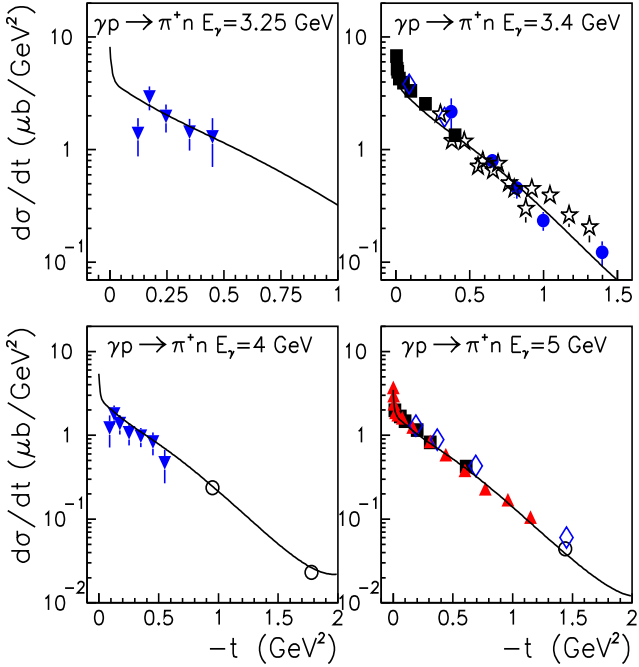,width=9.6cm,height=10.3cm}}
\vspace*{-5mm}
\caption{The $\gamma{p}{\to}\pi^+{n}$ differential cross section as a 
function of $-t$, the four-momentum transfer squared, at different
photon energies $E_\gamma$. The data are taken from Refs. \cite{Heide}
(filled squares), \cite{Joseph} (open diamonds), \cite{Boyarski1} (filled
triangles), \cite{Dowd} (inverse filled triangles), \cite{BarYam} (filled
circles), \cite{Elings} (open stars), and \cite{Anderson1,Anderson2} (open
circles). The solid lines show results of our model calculation based 
on the parameters listed in Table~\ref{tabp}.}
\label{gpion3}
\end{figure}
\begin{figure}[h]
\vspace*{-6mm}
\centerline{\hspace*{6mm}\psfig{file=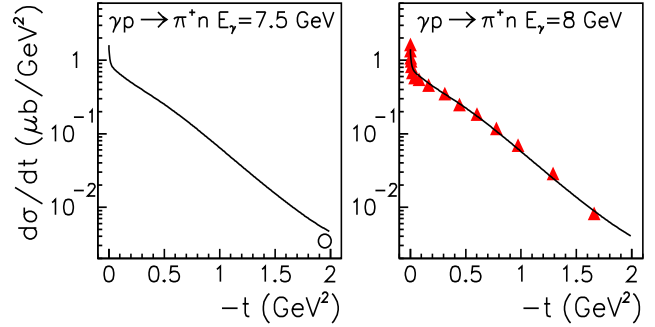,width=9.6cm,height=10.3cm}}
\vspace*{-49mm}
\caption{The $\gamma{p}{\to}\pi^+{n}$ differential cross section as a function 
of $-t$ at different
photon energies $E_\gamma$. The data are taken from Refs. \cite{Boyarski1}
(filled triangles) and \cite{Anderson1,Anderson2} (open circle). The solid
lines show results of our model calculation.}
\label{gpion15_r1}
\end{figure}

Finally, we compare our predictions with the most
recent data~\cite{Zhu,Zhu1} for differential cross sections at
$1.1{\leq}E_\gamma{\leq}5.5$~GeV, collected by the Hall~A Collaboration
at JLab. The main interesting feature of these data is that they cover a region
of fairly large momentum transfer -5.6${\leq}t{\leq}{-}0.4$ GeV$^2$ and thus
provide a window for examining the transition from non-perturbative QCD to
perturbative QCD. Also here we are guided by the aim to learn how the 
amplitudes generated from our Regge model could be used to investigate 
the reaction mechanisms relevant in this rather complex and exciting
transition region. However, the solution of this problem is beyond the
scope of this paper and will be postponed to future investigations.

\begin{figure}[b]
\vspace*{-9mm}
\centerline{\hspace*{6mm}\psfig{file=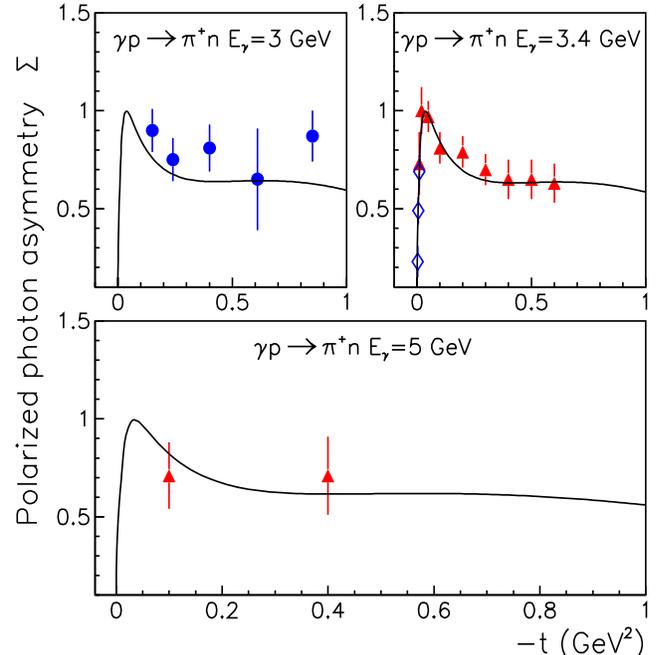,width=9.7cm,height=10.cm}}
\vspace*{-5mm}
\caption{Polarized photon asymmetry for the reaction $\gamma{p}{\to}\pi^+{n}$
as a function of $-t$ at different 
photon energies $E_\gamma$. The data are taken from Refs.  
\cite{Geweniger} (filled triangles), \cite{BarYam2} (filled 
circles) and \cite{Burfeindt} (open diamonds).
The solid lines are the results of our calculation.}
\label{gpion18_r1}
\end{figure}

\subsection{Results at {\boldmath$E_\gamma \geq 3$ GeV}}
Our results for the $\gamma{p}{\to}\pi^+{n}$ differential cross sections are 
presented in Figs.~\ref{gpion3} and \ref{gpion15_r1}. The model 
reproduces the data quite well.
Note that the differential cross sections increase sharply when approaching 
$t{=}0$ and can only be fitted by using  
the gauge invariant pion exchange term $F^{(\pi)}$ as described in 
subsection \ref{pionsec}. It will be interesting to see whether this 
particular feature can give us some clue about how the Regge model can be
connected with meson-exchange models. In the latter a gauge invariant 
pion-exchange, derived from phenomenological Lagrangians, is also 
a crucial ingredient in describing the charged pion photoproduction at
lower energies $\sqrt{s}{\leq}2$ GeV. In particular, it will be instructive to 
compare the multipole amplitudes in the transition region 
$\sqrt{s} \simeq 2-3$ GeV
where both models could be equally successful in describing the non-resonant
contribution around $t{=}0$. Our investigation on this issue will be
reported elsewhere.

\begin{figure}[t]
\vspace*{-6mm}
\centerline{\hspace*{5.mm}\psfig{file=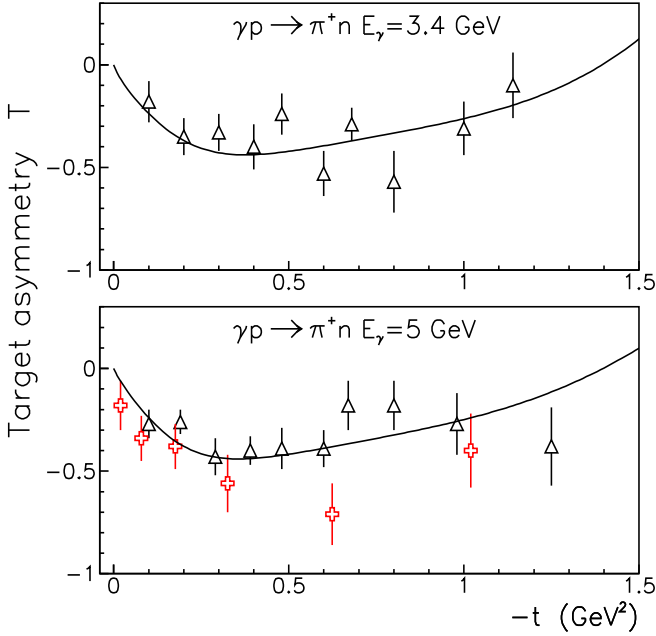,width=9.5cm,height=9.8cm}}
\vspace*{-5mm}
\caption{Target asymmetry $T$ for the reaction $\gamma{p}{\to}\pi^+{n}$
as a function of $-t$ at different 
photon energies $E_\gamma$. The data are taken from Refs.  
\cite{Genzel} (open triangles) and \cite{Morehouse} (open crosses).
The solid lines show results of our model calculation.}
\label{gpion17_r1}
\end{figure}

\begin{figure}[t]
\vspace*{-6.mm}
\centerline{\hspace*{4.5mm}\psfig{file=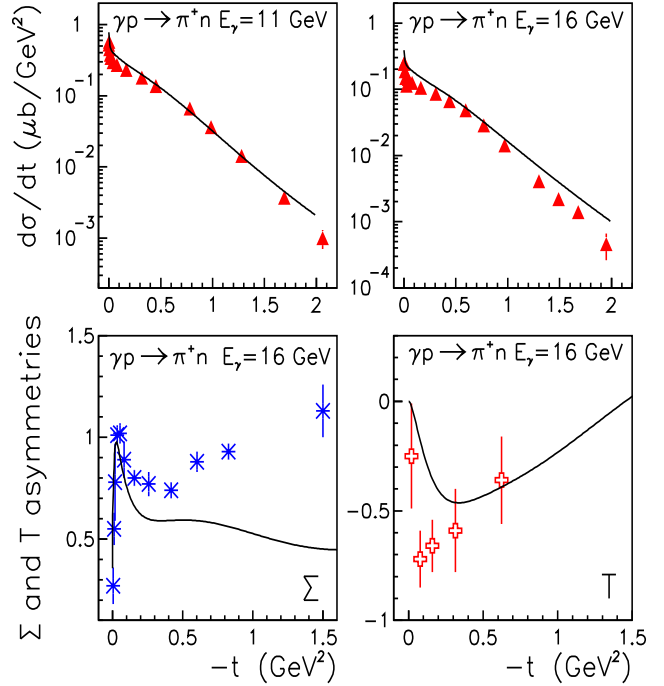,width=9.4cm,height=10.6cm}}
\vspace*{-5mm}
\caption{The $\gamma{p}{\to}\pi^+{n}$ differential cross
section (upper panel) and polarized photon, $\Sigma$ and target $T$
asymmetries as a function of $-t$ 
at the photon energies $E_\gamma$=11~GeV and 16~GeV. 
The data are taken from Refs. \cite{Boyarski1} (filled triangles),
\cite{Sherden} (asterisk) and \cite{Morehouse} (open crosses).
The solid lines show results of our model calculation.}
\label{gpion15_r2}
\end{figure}

\begin{figure}[t]
\vspace*{-5.mm}
\centerline{\hspace*{5mm}\psfig{file=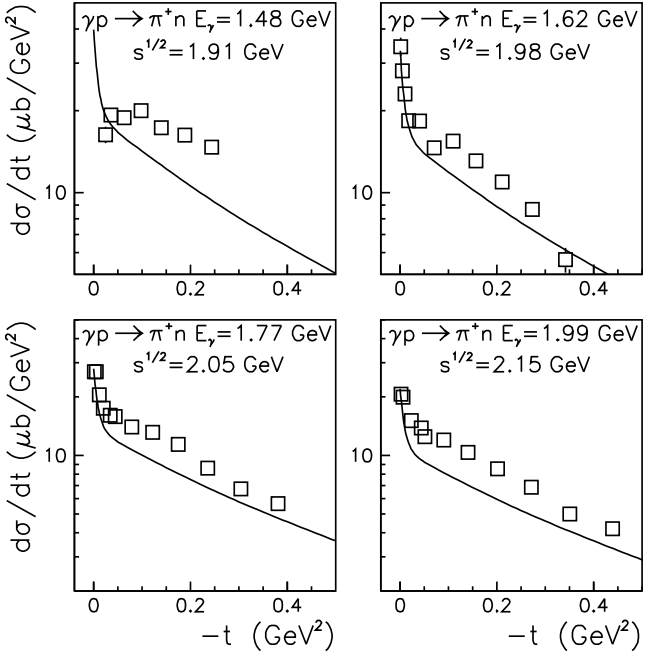,width=10.cm,height=10.1cm}}
\vspace*{-5mm}
\caption{The $\gamma{p}{\to}\pi^+{n}$ differential cross
section as a function of $-t$ at different
photon energies $E_\gamma$. Here $\sqrt{s}$  is the $\gamma{p}$ invariant
collision energy. The data are taken from Refs. \cite{Buschhorn1,Buschhorn2} 
(open squares). The solid lines show results of our model calculation.}
\label{gpion20}
\end{figure}

\begin{figure}[t]
\vspace*{-5mm}
\centerline{\hspace*{5mm}\psfig{file=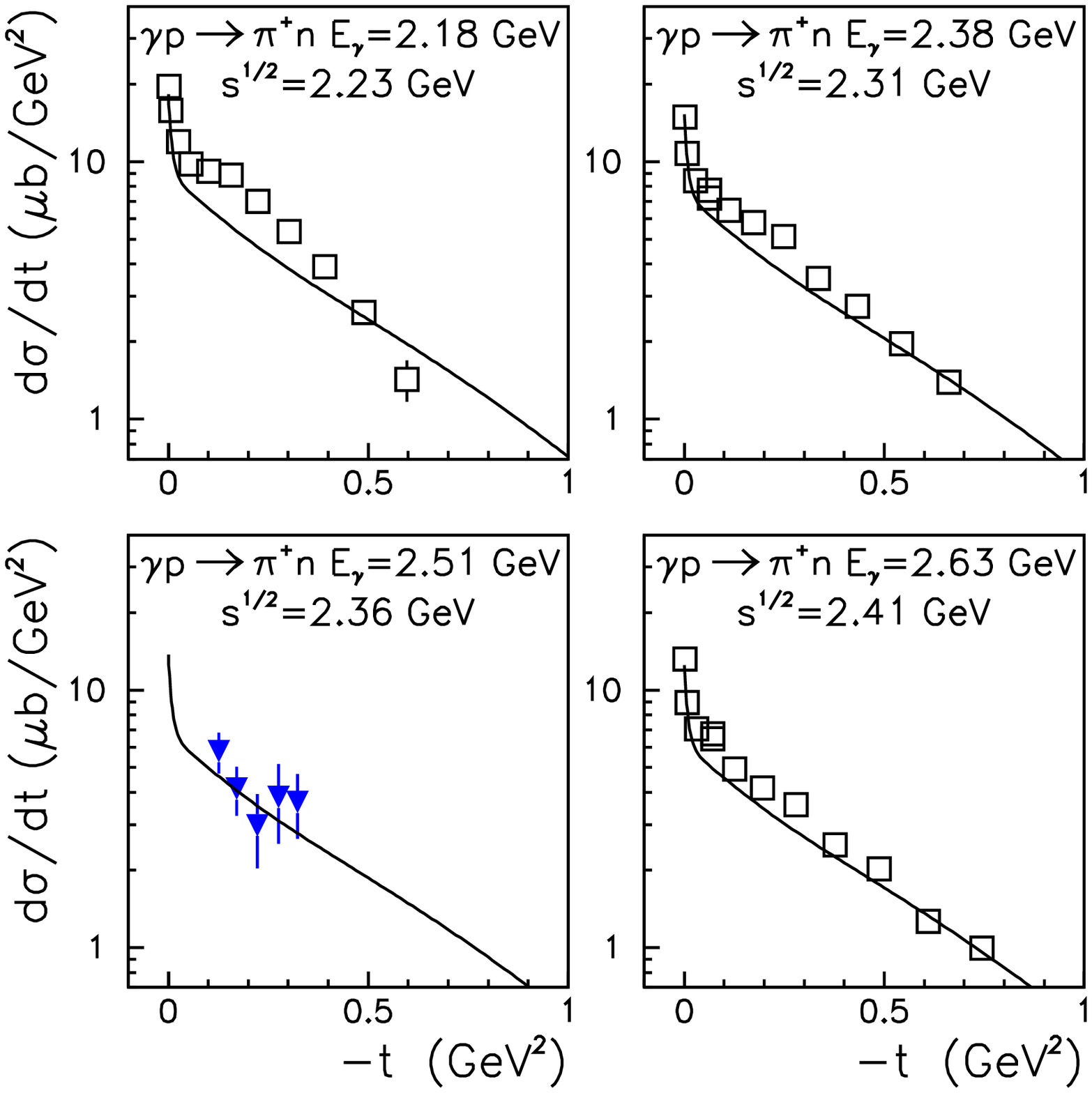,width=10.cm,height=10.1cm}}
\vspace*{-5mm}
\caption{The $\gamma{p}{\to}\pi^+{n}$ differential cross
section as a function of $-t$ at different
photon energies $E_\gamma$. The data are taken from Refs. \cite{Dowd}
(inverse close triangles) and \cite{Buschhorn1,Buschhorn2} (open squares). The 
solid lines show results of our model calculation.}
\label{gpion16}
\end{figure}

Fig.~\ref{gpion18_r1} presents results of our fits to the
data for the photon asymmetry $\Sigma$. The photon asymmetry 
$\Sigma$ is defined by   
\begin{eqnarray}
\Sigma = \frac{d\sigma_\perp - d\sigma_\parallel}{d\sigma_\perp +
d\sigma_\parallel} \ ,
\label{paral}
\end{eqnarray}
where $d\sigma_\perp$ ($d\sigma_\parallel$) is the cross section
from measurements with photons polarized in the direction perpendicular
(parallel) to the $\gamma$-$\pi$ scattering plane. 

\begin{figure}[h]
\vspace*{-6mm}
\centerline{\hspace*{6mm}\psfig{file=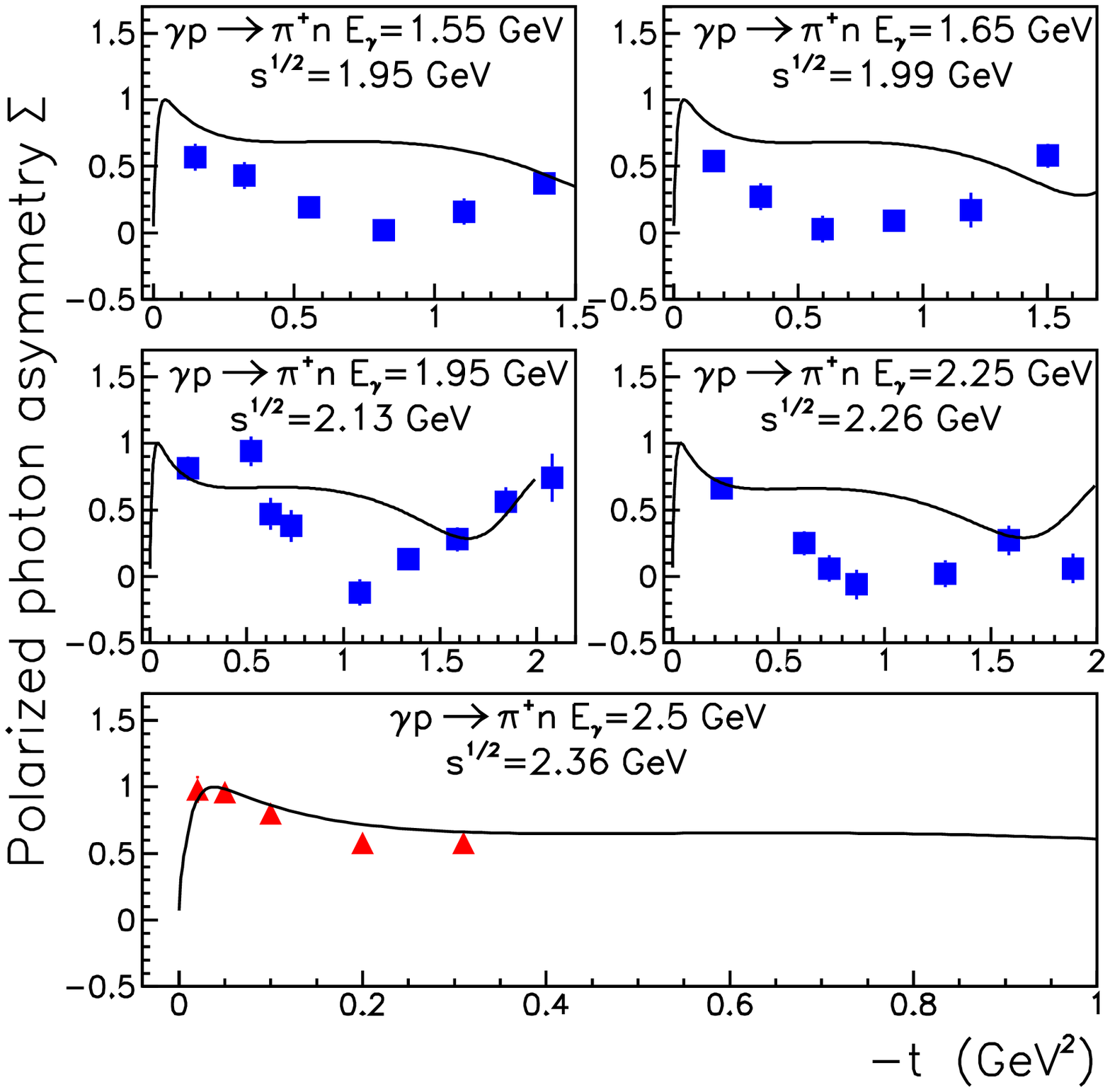,width=9.8cm,height=11.5cm}}
\vspace*{-6mm}
\caption{Polarized photon asymmetry from
$\gamma{p}{\to}\pi^+{n}$ reaction as a function of $-t$ 
at different photon energies. The data are taken from
Refs. \cite{Bussey2} (squares) and \cite{Geweniger} (filled triangles). 
The solid lines show results of our model calculation.}
\label{gpion17b}
\end{figure}

Within the Regge model the 
data on $\Sigma$ can be used~\cite{Stichel,Ravndal,Bajpai} to separate
the contributions from natural and unnatural parity exchanges.
According to Ref.~\cite{Stichel} the differential cross section
$d\sigma_\perp$ ($d\sigma_\parallel$) is due to unnatural (natural) parity 
exchanges. Thus, a large and positive $\Sigma$ at forward angles, seen in the
left panel of Fig.~\ref{gpion18_r1}, indicates the dominance of unnatural 
parity exchanges and 
can be described by the pion-exchange mechanism~\cite{BarYam2,Burfeindt}. 
The data show that $\Sigma$ is positive 
and almost constant, suggesting that $d\sigma_\perp{>}d\sigma_\parallel$ over 
the whole considered range of $t$. Hence, the photoproduction of $\pi^+$
photoproduction is indeed dominated by unnatural parity exchanges in the 
kinematic region considered. 

Our fit to the data on the target asymmetry $T$ is shown in
Fig.~\ref{gpion17_r1}.  
The data were obtained~\cite{Genzel,Morehouse} using a buthanol frozen spin target. 
Within the experimental uncertainties the data are well reproduced by our calculation.
To test the constructed model, we first compare our predictions with the data available at 
higher energies, namely at $E_\gamma$=11~GeV and 16~GeV. Note that these data were 
not included in our global fits. The upper panel of Fig.~\ref{gpion15_r2} shows 
the differential cross section for $\gamma{p}{\to}\pi^+{n}$, which is well
reproduced by the model calculation. 
The lower panel of Fig.~\ref{gpion15_r2} displays results for the polarized photon 
asymmetry $\Sigma$ and the target asymmetry $T$ for $E_\gamma$=16~GeV. Here deficiencies
of the model are apparent. 

To remove the remaining discrepancies, specifically in the polarization observables, one 
may have to include the $F_4$ contribution which has been neglected in the present fit,
as discussed in subsection \ref{f4sec}. Indeed, the $F_4$ contribution is primarily 
sensitive to the difference between the recoil ($R$) and target ($T$)
asymmetries, as given by Eq.~(\ref{bound1}). Unfortunately, there is no experimental
information available for $R$. Apparently more data on $\Sigma$, $T$ and $R$ as well 
as other polarization observables are needed for making further progress.

\subsection{Predictions at lower energies}
As was stressed in many
studies~\cite{Irving,Worden,Henyey}, the Regge theory 
is phenomenological in nature.
There is no solid theoretical derivation that allows us 
to establish explicitly the ranges of $t$ and $s$ where 
this formalism is applicable. Nevertheless,
following the usual arguments based on the
analytic properties of the scattering amplitudes in the complex
angular momentum plane, it is reasonable to assume that
the Regge model constructed above is valid for describing quantitatively
the exchange mechanisms down to energies of around $E_\gamma{\simeq
}3$~GeV which corresponds to $\sqrt{s}{\simeq}2.55 $ GeV.
Since there are several well-identified nucleon resonances~\cite{Yao} 
in the energy range up to the range of $\sqrt{s}{\simeq}2.6 $ GeV, 
identified in partial wave
analyses~\cite{Hoehler1,Hoehler2,Koch,Cutkovsky,Manley,Penner,Arndt,Batinic} of
pion-nucleon scattering, 
we expect that deviations of our predictions from the data will start to 
show up for energies from $E_\gamma{\simeq}3 $ GeV downwards and it is 
obvious that those discrepancies could be a signal for possible 
contributions from nucleon resonances. Thus, 
our specific interest here is to examine carefully this transition energy 
region and to single out those observables which can be
used most effectively to establish the presence of resonances 
or even to extract nucleon resonance parameters.

\begin{figure}[t]
\vspace*{-6.mm}
\centerline{\hspace*{6mm}\psfig{file=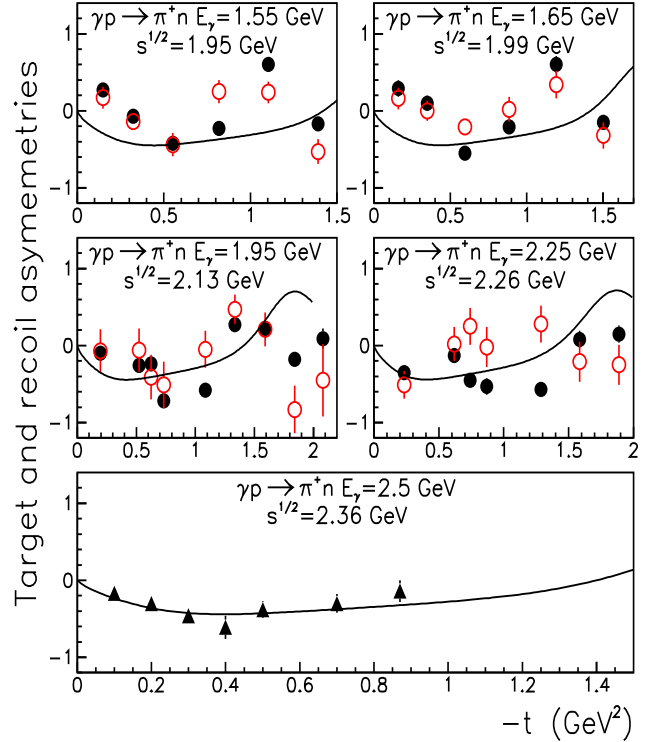,width=9.8cm,height=11.5cm}}
\vspace*{-6mm}
\caption{Target ($T$) (filled circles and triangles), and recoil asymmetry 
($R$) (open circles) for
$\gamma{p}{\to}\pi^+{n}$ as a function of $-t$ at different photon energies
$E_\gamma$. The data are taken from Refs. \cite{Bussey2} (filled and open circles)
and \cite{Genzel} (triangles). The solid lines are our result.
}
\label{gpion17a}
\end{figure}

The solid lines in Figs.~\ref{gpion20},\ref{gpion16} show our predictions 
for the $\gamma{p}{\to}\pi^+{n}$ differential cross sections at 
$1.48{\le}E_\gamma{\le}2.63$~GeV in comparison with the data.
Here we also indicate the corresponding $\gamma{p}$ invariant
mass $\sqrt{s}$. We see from Fig.~\ref{gpion16} that our predictions are in 
reasonable agreement with the experimental results~\cite{Dowd,Buschhorn1,Buschhorn2}
down to $E_\gamma$=2.38 GeV, which corresponds to an invariant mass of 
$\sqrt{s}{\simeq}2.31$~GeV. At those energies there is not much room for 
additional contributions within the $t$ range covered by the 
experiments. As seen from Figs.~\ref{gpion20} and \ref{gpion16}, our predictions 
start to deviate more systematically from the data below $E_\gamma =2.18$ GeV 
or $\sqrt{s}=2.23$ GeV.

Our predictions for the photon asymmetry $\Sigma$ are presented in Fig.~\ref{gpion17b}.
Here we see very large differences between our predictions (solid lines) and the data
for photon energies $E_\gamma \le$ 2.25~GeV. On the other hand, the model
is in good agreement with data on $\Sigma$ at the photon energy 
$E_\gamma$=2.5~GeV, which corresponds to $\sqrt{s}{=}$2.36~GeV. But here one should
keep in mind that the data cover only a very small range of $t$.  
Since the polarized photon asymmetry varies substantially as a function of the 
four-momentum transfer squared within the considered range 1.95${\le}\sqrt{s}{<}$2.36~GeV, 
one might consider this as an indication for the excitation of baryonic resonances.

Fig.~\ref{gpion17a} presents data on the target polarization $T$ (filled circles 
and triangles) and the recoil polarization $R$ (open circles) together with 
the model results.
Please recall that in our model we assume $F_4{=}0$ and, hence, the 
predictions for these two observables are the same, cf. Eqs.~(\ref{obs3}) and 
(\ref{obs4}). Thus, there is only one (solid) line in each panel of Fig.~\ref{gpion17a}.
There are some deviations of our model result from the data
at photon energies below 2.25~GeV. However, the accuracy of the data is not 
sufficient to draw further and more concrete conclusions.
Indeed, it looks as if both $R$ and $T$ oscillate around the value zero.
It is interesting to note that the data in Fig.~\ref{gpion17a} suggest that $T{\simeq}R$ 
within the experimental uncertainties. Thus, our assumption that $F_4=0$
is in line with the experimental evidence. Nevertheless, more precise data on these 
two observables would be rather useful for drawing more definite conclusions on $F_4$.

\subsection{Comparison with the JLab data}
The most recent data on charged meson photoproduction were obtained by
the Hall A Collaboration~\cite{Zhu,Zhu1} at JLab. 
These data cover a wide range of photon energies
(1.1${\leq}E_\gamma{\leq}5.53$ GeV) and squared four-momentum transfers
(0.4${\leq}-t{\leq}$5.6 GeV$^2$).

\begin{figure}[t]
\begin{center}
\vspace*{-6mm}
\hspace*{-4mm}\psfig{file=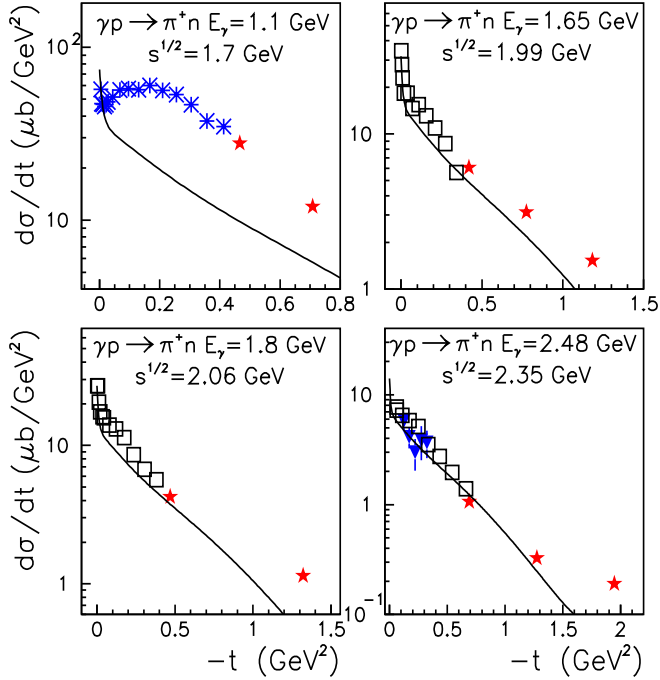,width=9.9cm,height=10.5cm}\\
\vspace*{-5mm}
\caption{The $\gamma{p}{\to}\pi^+{n}$ differential cross
section as a function of $-t$ at different
photon energies $E_\gamma$. The data are taken from Refs. \cite{Dowd} (inverse
close triangles), \cite{Buschhorn1,Buschhorn2} (open squares) and \cite{Ecklund}
(asterisk). The stars are the experimental results from the JLab Hall A 
Collaboration~\cite{Zhu}. The solid lines show results of our model calculation.}
\label{gpion21}
\end{center}
\vspace*{-2mm}
\end{figure}

We first consider the data at low energies, $E_\gamma{\leq}2.48$ GeV.
Fig.~\ref{gpion21} presents differential cross sections as a function of 
the squared four-momentum transfer collected from different 
experiments~\cite{Dowd,Buschhorn1,Buschhorn2,Ecklund,Zhu}. Here the JLab 
data~\cite{Zhu} correspond to the stars and they are consistent with previous 
measurements. (Note that the other data were taken, in general, at 
slightly different energies, cf. Figs.~\ref{gpion3},
\ref{gpion15_r1}, \ref{gpion20}, \ref{gpion16}.)
The solid lines are the results of our model calculation. They
are in line with the JLab data points for $E_\gamma{\ge}$1.65~GeV and 
for small ${-}t$, but deviate from the data at -$t$ around or above 1~GeV. 
The large discrepancy at $E_\gamma{=}$1.1~GeV or 
$\sqrt{s}$=1.7~GeV is to be expected because in this energy region there should 
be additional contributions from well established resonances. 

\begin{figure}[t]
\begin{center}
\vspace*{-6.mm}
\hspace*{-4mm}\psfig{file=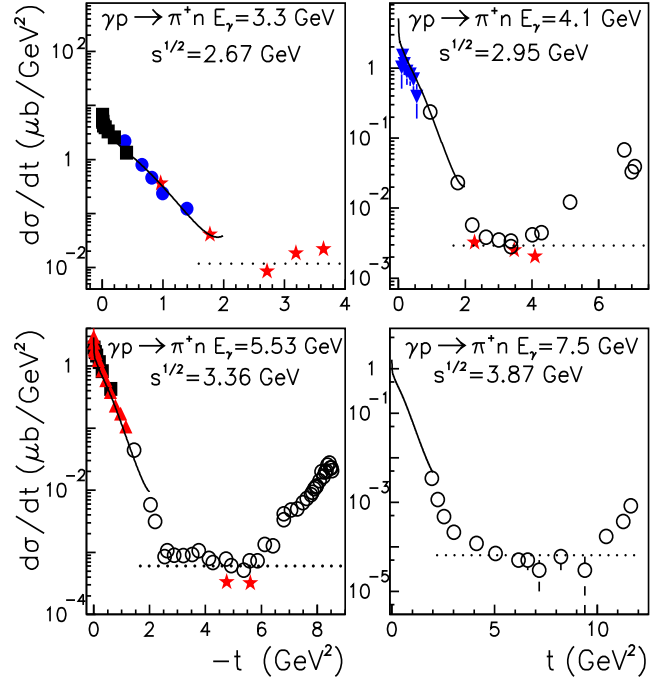,width=9.9cm,height=10.5cm}
\vspace*{-9mm}
\caption{The $\gamma{p}{\to}\pi^+{n}$ differential cross
section as a function of $-t$ at different
photon energies $E_\gamma$. The data are taken from Refs. \cite{Heide}
(filled squares), \cite{Boyarski1} (filled triangles), \cite{BarYam}
(filled circles) and \cite{Anderson1,Anderson2} (open circles).
The stars are the
experimental results from the JLab Hall A Collaboration~\cite{Zhu}.
The solid lines show our results based on the parameters listed in
Table~\ref{tabp}. The dotted lines are results obtained 
with Eq.~(\ref{count}).}
\label{gpion24}
\end{center}
\end{figure}

In Fig.~\ref{gpion24} we compare our predictions at $E_\gamma{\geq}3.0$ GeV
with the JLab data (stars) and with all other available data.
Note that the older data shown by open circles in the figure for 
$E_\gamma$=4.1~GeV and 5.53~GeV are actually from measurements 
at $E_\gamma$=4.0~GeV and 5.0~GeV, respectively. However, these small energy 
differences are not important for our discussions here.
Obviously only two of the JLab data points at the photon 
energy $E_\gamma$=3.3~GeV are in the $-t\leq 2$ GeV$^2$ region. These are 
well described by our model prediction. Furthermore, they are also in 
good agreement with data from earlier measurements~\cite{Heide,BarYam}.
The other JLab data as well as all older experimental results 
for the higher $|t|$ region are simply beyond the applicability of our model.

Above $|t|{\simeq}2$~GeV the data show first an almost $t$ independent behavior and 
then increase sharply as $-t$ approaches its maximum value.
The largest -$t$ value corresponds to the smallest value of
$|u|$, which is related to $s$ and $t$ by $s{+}t{+}u{=}2m_N^2{+}m_\pi^2$.
It is known that the reaction mechanism at small $|u|$ and at small $|t|$  
involves different exchanges. The reaction at small $|t|$ is dominated by the 
meson poles and cuts included in the Regge model constructed in this work. On 
the other hand, the rising cross sections at small $|u|$ (large $|t|$) observed 
in Fig.~\ref{gpion24} are due to the exchange of baryon resonances.

In the central region 2${\leq}{-}t{\leq}$5 GeV$^2$ of Fig.~\ref{gpion24},
both $|t|$ and $|u|$ are large and hence the contributions from
$t$- and $u$-channel exchanges become very small.
The main feature of the cross sections in this middle region is that
they are almost independent of $t$ and hence are very unlikely due to
nucleon resonances with reasonably narrow widths, {\it i.e.} with widths $\leq$ 300 MeV.
The most plausible interpretation can be found from the point of view of 
perturbative QCD. The essential idea is that at large momentum transfer
the basic interactions must be directly due to the quarks in the nucleon. 
As was proposed in Refs.~\cite{Brodsky,Matveev} the energy
dependence of the reaction cross sections for this case is driven by the total number of
elementary fields in the initial ($n_i$) and final ($n_f$) states. Following dimensional 
counting for the invariant amplitude~$\cal{M}$~\cite{Byckling} the energy dependence of 
the differential cross section of the $n_ i{\to}n_f$ transition is given as
\begin{eqnarray}
\frac{d\sigma}{dt}{=}\frac{|{\cal
M}|^2 F(t) }{16\pi(s{-}m_N^2)^2} \stackrel{^{m_N^2{\ll}s}}{=}
&&\frac{s^{{-}(n_i{-}2){-}(n_f{-}2)}F(t)}{16\pi s^2}\nonumber \\
\propto
&&s^{-7} F(t) \ ,
\label{count}
\end{eqnarray}
since for single pion photoproduction $n_i{=}4$ and $n_f{=}5$. Here $F(t)$ does
not depend on $s$ but accounts for the $t$-dependence of the hadronic wave
functions and partonic scattering.

In order to see whether the data shown in Fig.\ref{gpion24} follow the 
dimensional counting rule (also called the quark counting rule), we 
normalize the expression for $d\sigma/dt$ in Eq.~(\ref{count}) 
with $F(t)=1$ to the data at $t{=}5$~GeV$^2$ and at $E_\gamma$=7.5 GeV, 
{\it i.e.} at the highest of the considered photon energies, and then use
Eq.~(\ref{count}) to predict the cross sections at other energies.
These predictions are shown by dotted lines in Fig.~\ref{gpion24} and
agree remarkably well with the data at all considered energies.
It appears that the dimensional counting rule, as given in
Eq.~(\ref{count}), is fulfilled very well. Such a conclusion was drawn also
in Refs.~\cite{Zhu,Zhu1} by analyzing the JLab data alone.
It is an outstanding challenge to understand this smooth $t-$dependence.
One possibility is to explore more rigorously the handbag 
mechanism~\cite{Huang1,Huang2}, which yields a reasonable description of
the $\pi^+{/}\pi^-$ ratio.

\section{Results for {\boldmath$\gamma n \rightarrow \pi^-p$}}

The strategy for the analysis of negative pion photoproduction is similar to 
that described in Section~\ref{secstra} for positive pions.
Some general information on the data on $\gamma{n}{\to}\pi^-p$ included
in our global fit is listed in the Tables~\ref{tab:data+1} and \ref{tab:data+2}. 

The measurement of the $\gamma{n}{\to}\pi^-{p}$ reaction can be only done
with a deuteron target.
The extraction of data for negative pion photoproduction from the deuteron reaction
is based on the so-called spectator model, {\it i. e.} the single scattering impulse
approximation. Thereby, it is assumed that the proton of the deuteron is the
spectator and its role in the $\gamma{n}$ interaction is only due to the Fermi
motion of the bound neutron. 
This is, in principle, a reliable  
method~\cite{Meyer,Duncan,Hann,Calen,Zlomanczuk,Moskal,SibirtsevK,TOF}
as long as one measures the momentum distribution of the proton and one takes
only those events which fulfill the spectator condition, {\it i. e.} those events 
where the 
proton momentum is smaller than the momentum of the neutron. However, in
practice often the spectator proton and the final neutron are not even
identified. In that case one might expect~\cite{SibirtsevK} some 
discrepancies between the model calculations and the data as well as between
different measurements. Some important details of the deuteron experiments will be 
given in the following in order to discuss possible reasons 
for the observed discrepancies.

\begin{table}[t]
\begin{center}
\caption{The $\gamma{n}{\to}\pi^-p$ data on differential cross section
analyzed in the present paper. }
\label{tab:data+1}
\begin{tabular}{|r|r|r|r|c|}
\hline\noalign{\smallskip}
\multicolumn{5}{|c|}{ Differential cross section,  $d\sigma{/}dt$ } \\
\hline\noalign{\smallskip}
$E_\gamma$ \,\, & $\sqrt{s}$\,\, & $-t_{min}$  & $-t_{max}$  & Reference  \\ 
(GeV) & (GeV) & (GeV$^2$) & (GeV$^2$) &  \\ 
\noalign{\smallskip}\hline\noalign{\smallskip}
1.1 & 1.7  & 0.41 & 1.11 & \cite{Sternemann}\\
1.1 & 1.7  & 0.41 & 1.11 & \cite{Scheffler} \\
1.1 & 1.7  & 4.2${\times}10^{-3}$  & 1.37 & \cite{Sternemann}\\
1.1 & 1.7  & 0.25  & 0.71 & \cite{Zhu}\\
1.65 & 1.99  & 0.72 & 1.49 & \cite{Sternemann}\\
1.65 & 1.99  & 0.42  & 1.19 & \cite{Zhu}\\
1.8& 2.06  & 0.80 & 2.10 & \cite{Sternemann}\\
1.8& 2.06  & 0.19 & 2.64 & \cite{Benz}\\
1.8& 2.06  & 0.48 & 1.33 & \cite{Zhu}\\
2.48& 2.35  & 0.69 & 1.94 & \cite{Zhu}\\
3.0 & 2.55 & 0.15 & 1.16 & \cite{BarYam,BarYam2} \\
3.32 & 2.67 & 0.96 & 3.64 & \cite{Zhu}\\
3.4 & 2.69 & 0.37 & 1.39 & \cite{BarYam,BarYam2} \\
3.4 & 2.69 & 3.0${\times}10^{-3}$ & 0.4 & \cite{Heide} \\
4.15 & 2.95  & 1.25 & 3.47 & \cite{Zhu}\\
5.0 & 3.2 & 6.8${\times}10^{-3}$ & 0.53 & \cite{Heide} \\
5.53 & 3.36  & 3.18 & 4.73 & \cite{Zhu}\\
8.0 & 3.99 & 9.9${\times}10^{-3}$ & 0.89 & \cite{Boyarski2} \\
\noalign{\smallskip}\hline\noalign{\smallskip}
\end{tabular}
\end{center}
\end{table}

\begin{table}[h]
\begin{center}
\caption{The $\gamma{n}{\to}\pi^-p$ data on the polarized photon asymmetry
$\Sigma$ (denoted formerly as $A$~\cite{Worden}) analyzed in the present
paper.}
\label{tab:data+2}
\begin{tabular}{|r|r|r|r|c|}
\hline\noalign{\smallskip}
\multicolumn{5}{|c|}{ Polarized photon asymmetry $\Sigma$ } \\
\hline\noalign{\smallskip}
$E_\gamma$ \,\, & $\sqrt{s}$\,\, & $-t_{min}$  & $-t_{max}$  & Reference  \\ 
(GeV) & (GeV) & (GeV$^2$) & (GeV$^2$) &  \\ 
\noalign{\smallskip}\hline\noalign{\smallskip}
3.0 & 2.55 &0.15 & 1.16 & \cite{BarYam2} \\ 
3.4 & 2.69 &0.05 & 0.6 & \cite{Burfeindt} \\ 
16.0 & 5.56 &5.5$\times$10$^{-3}$  & 1.19 & \cite{Sherden} \\ 
\noalign{\smallskip}\hline\noalign{\smallskip}
\end{tabular}
\end{center}
\end{table}

\begin{figure}[b]
\vspace*{-8mm}
\centerline{\hspace*{4mm}\psfig{file=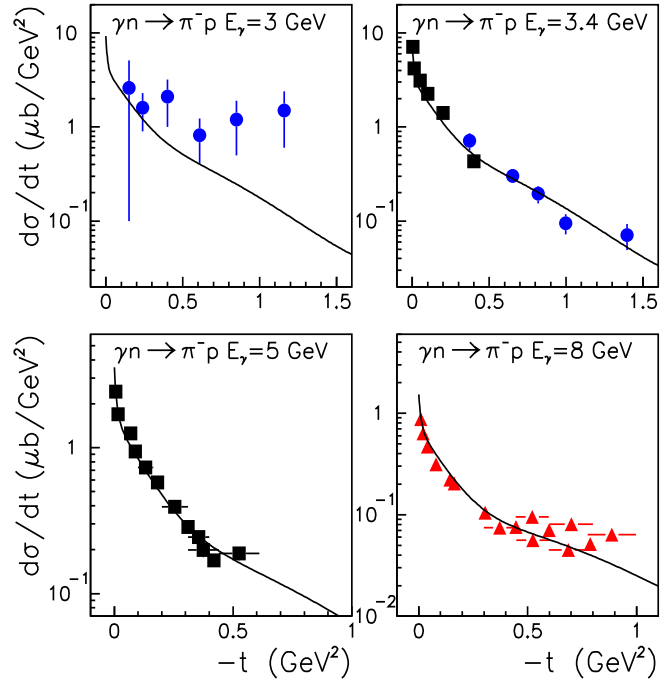,width=10.cm,height=10.5cm}}
\vspace*{-5mm}
\caption{The $\gamma{n}{\to}\pi^-{p}$ differential cross
section as a function of $-t$ at different
photon energies $E_\gamma$. The data are taken from Refs.
\cite{Heide} (filled squares), \cite{BarYam,BarYam2} (filled circles), 
and \cite{Boyarski2} (filled triangles). The solid lines show results of
our model calculation.}
\label{gpion14}
\end{figure}

\subsection{Results at {\boldmath$E_\gamma \geq 3$ GeV}}
Fig.~\ref{gpion14} shows differential cross sections for
$\gamma{n}{\to}\pi^-{p}$  
measured~\cite{Heide,BarYam,Boyarski2} at different photon energies together 
with results of our model calculation. Indeed, these are practically all 
$\pi^-$ photoproduction data for $E_\gamma \geq 3$ GeV that are available 
in the literature. 

Let us first provide some details on the above experiments which will be 
useful later in discussing the observed discrepancies between the older 
measurements and the most recent results from JLab 
reported by the Hall A Collaboration~\cite{Zhu,Zhu1}. 
In the experiment of Ref.~\cite{Heide} $\gamma{d}{\to}\pi^-2p$,  
$\gamma{d}{\to}\pi^+2n$, and in addition $\gamma{p}{\to}\pi^+n$ were studied 
in order to check the validity of the spectator model. The experiment was performed
at DESY with a bremstrahlung beam and by detecting only the pions with a magnetic spectrometer.
At small $-t$ the relation between the photon energy and the pion momentum is almost identical 
to the one for photoproduction on a free nucleon. Thus by measuring the pion momentum at a 
given angle one can reconstruct the photon energy. The Fermi motion in the deuteron results 
in an uncertainty of $\pm$100~MeV in the invariant mass energy of the final system. Furthermore,
utilizing simulation calculations of the reaction based on the Hulth\'en deuteron wave function 
it was found that the computed momentum spectrum of the pions is in good agreement with 
the measured one. The energetic 
separation between single and multiple pion photoproduction was good enough to avoid di-pion 
contamination. Single pion photoproduction was studied by using the photon energy interval of 200~MeV 
around $E_\gamma$=3.4~GeV and 5~GeV.

\begin{figure}[t]
\vspace*{-6.5mm}
\centerline{\hspace*{4mm}\psfig{file=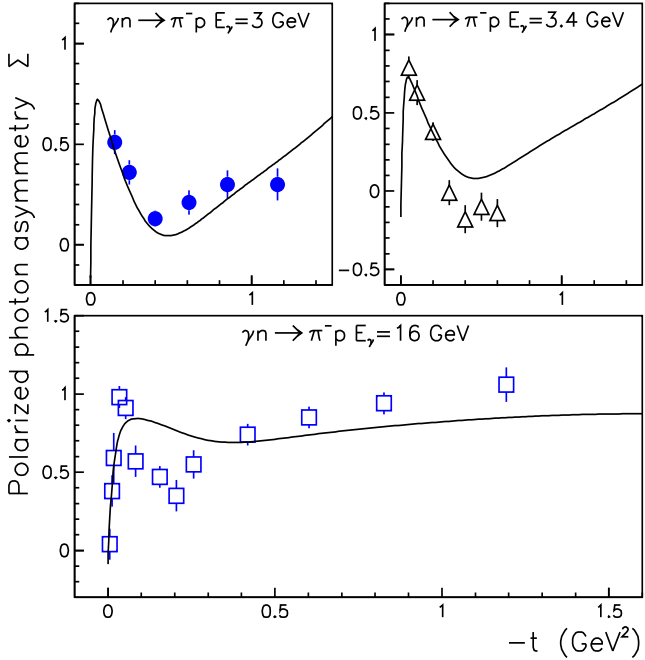,width=10.cm,height=10.2cm}}
\vspace*{-5mm}
\caption{Polarized photon asymmetry from $\gamma{n}{\to}\pi^-{p}$
reaction as a function of $-t$ at different 
photon energies, $E_\gamma$. The data are taken from Refs.  
\cite{BarYam2} (filled circles), \cite{Sherden} (open squares) and 
\cite{Burfeindt} (open triangles). The solid lines show results of our model
calculation.}
\label{gpion19}
\end{figure}

We should also mention that in Ref.~\cite{Heide} it was observed 
that at $|t|{\ge}$0.3~GeV$^2$ the differential cross 
sections for $\pi^+$-meson photoproduction on deuterium and hydrogen are almost
identical, 
while at smaller momentum transfers they differ substantially, {\it i. e.} up to
a 
factor of $\simeq$2. That was qualitatively understood from spin and isospin
restrictions of the 
spectator model~\cite{Baldin}. A similar suppression 
of the $\pi^+$-meson yield from deuterium at small angles was observed in
lower-energy 
experiments~\cite{Neugebauer}. Furthermore, the $\pi^-{/}\pi^+$ ratio was
evaluated
under the assumption that the corrections for the $\gamma{d}{\to}\pi^-2p$
and $\gamma{d}{\to}\pi^+2n$ reactions are the same.

The circles in Fig.~\ref{gpion14} are data taken~\cite{BarYam} at the Cambridge 
Electron Accelerator. The results at $E_\gamma$=3.4~GeV are
published~\cite{BarYam}, 
while the data for the differential cross section at $E_\gamma$=3~GeV, mentioned
in 
Ref.~\cite{BarYam2}, are available from the Durham Data Base~\cite {Durham}.
The $\gamma{d}{\to}\pi^-2p$, $\gamma{d}{\to}\pi^+2n$ 
and $\gamma{p}{\to}\pi^+n$ reactions were studied by detecting only the
$\pi^-$-meson. 
The reconstruction procedure for the reaction is almost identical to that
applied in Ref.~\cite{Heide}. The energy of the incident photon was 
determined by a subtraction method and could be evaluated to an 
accuracy of $\pm$60~MeV in the considered range from 3 to 3.7~GeV 
(explored in the search for the $N^\ast(2645)$ baryon). Under the assumption 
that 
the spectator nucleon is at rest the missing mass for an interacting 
nucleon was reconstructed in order to separate single pion photoproduction 
from multiple pion contributions. 

Our calculation reproduces the $\pi^-$-meson data at $E_\gamma$=3.4 GeV 
rather well. The differential cross section at $E_\gamma$=3~GeV is described 
qualitatively. But it looks as if some additional contribution is required 
for the range of $-t{\ge}0.4$ GeV$^2$, say, though  
one should keep in mind that the data at $E_\gamma$=3~GeV 
are afflicted by fairly large errors. 
In this context we want to recall that we reasonably reproduce the differential 
cross section and polarization data for positive pion photoproduction available around 
$E_\gamma$=3~GeV, cf. Figs.~\ref{gpion3} and \ref{gpion18_r1}.

One could speculate that this deviation of the model result from the data 
is a signal for an excited baryon with mass around 2.55~GeV. For instance, the 
$N^\ast(2645)$ resonance was observed in pion-nuclear interactions~\cite{Wahlig,Citron} 
but was not detected in the photoproduction of positive and neutral pions.
If the baryon is a member of a $U$-spin multiplet with $U$=3/2 it could not be 
excited in the interaction of photons with protons because the photon is considered 
to be $U{=}0$. In case of a neutron target both $n$ and 
the neutral $N^{\ast}$ have $U{=}1$ and the corresponding excitation is 
allowed~\cite{BarYam,BarYam2}. It is worth mentioning that there is no obvious
evidence for the presence of such a resonance in the polarized photon asymmetry 
shown in Fig.~\ref{gpion19}.  

The data at $E_\gamma$=8~GeV were measured~\cite{Boyarski2} at the Stanford
Linear Accelerator.
Again only pions were detected and the reaction was reconstructed by measuring the
pion momentum distribution resulting from photons near the bremsstrahlung dip. It was 
emphasized that such a reconstruction of single pion photoproduction is quite reasonable at small
$-t$ but becomes impractical at $|t|{\ge}2$~GeV, unless the other final state particles are also
detected. To test the spectator mechanism, $\gamma{d}{\to}\pi^+2n$ as well as
$\gamma{p}{\to}\pi^+n$ reactions were studied. It was found~\cite{Boyarski2} that at 
$|t|{<}$0.5~GeV$^2$ the differential cross sections for $\pi^+$-meson photoproduction on 
deuterium and hydrogen differ up to a factor of around 8. The reasons for such a discrepancy 
were investigated in detail and it was argued that the Pauli exclusion principle explains 
completely the observed effect. The relevant corrections were done for presenting the 
$\pi^-$-meson photoproduction data. Fig.~\ref{gpion14} clearly proofs that we perfectly 
reproduce the $\pi^-$-meson photoproduction differential cross section at $E_\gamma$=8~GeV.

\begin{figure}[t]
\vspace*{-6mm}
\centerline{\hspace*{4mm}\psfig{file=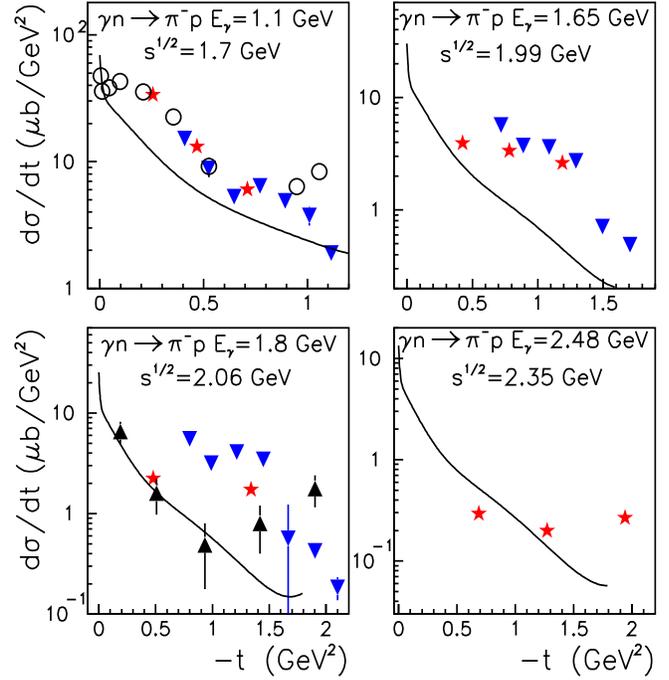,width=10.cm,height=10.5cm}}
\vspace*{-5mm}
\caption{The $\gamma{n}{\to}\pi^-{p}$ differential cross
section as a function of $-t$ at different
photon energies $E_\gamma$. The data are taken from Refs.  \cite{Scheffler}
(open circles), \cite{Sternemann} (filled inverse triangles) and \cite{Benz}
(filled triangles). The stars are the
experimental results from the JLab Hall A Collaboration~\cite{Zhu}.
The solid lines show our results based on the parameters listed in
Table~\ref{tabp}. 
}
\label{gpion22}
\end{figure}

Finally, in Fig.~\ref{gpion19} data on the polarized photon asymmetry for the reaction 
$\gamma{n}{\to}\pi^-p$ \cite{BarYam2,Sherden,Burfeindt} at photon energies 
3, 3.4 and 16~GeV are presented. 
In these experiments the reaction was reconstructed similar to the procedures
described above. The model calculation describes the experimental results well -- with 
exception of some data points. 

\subsection{Comparison with the JLab data}
Differential cross sections for $\gamma{n}{\to}\pi^-p$ 
at photon energies between 
1.1~GeV and 5.5~GeV were reported~\cite{Zhu,Zhu1} recently 
by the Hall A Collaboration at JLab. 
Although most of the data were obtained for large $|t|$, 
at some photon energies the measurements
extend to the region of $|t|{<}$2~GeV$^2$ and can be directly 
compared with our calculation.

We want to emphasize that this experiment with a deuterium target 
has some significant advantages 
as compared to the other measurements discussed in 
the previous subsection. In particular, both 
the $\pi^-$-meson and the proton were
detected in coincidence. Based on two-body kinematics, 
the incident photon energy was reconstructed.
That allows one to reconstruct the spectator momentum distribution 
which was found to be in good 
agreement with the Argonne, Paris and Bonn deuteron 
wave functions at momenta below 400~MeV/c.

\begin{figure}[b]
\vspace*{-6.5mm}
\centerline{\hspace*{4mm}\psfig{file=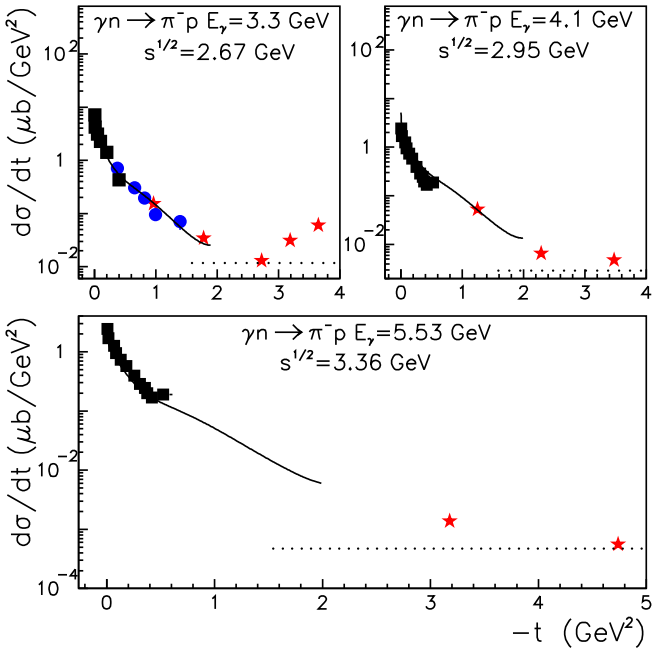,width=10.cm,height=10.cm}}
\vspace*{-5mm}
\caption{The $\gamma{n}{\to}\pi^-{p}$ differential cross
section as a function of $-t$ at different
photon  energies, $E_\gamma$. The data are taken from Refs. \cite{BarYam}
(filled circles) and \cite{Heide} (filled squares). The stars are the
experimental results from the JLab Hall A Collaboration~\cite{Zhu}.
The solid lines show our results based on the parameters listed in
Table~\ref{tabp}. The
dotted line shows result obtained by Eq.~(\ref{count}) and normalized to the
$\gamma{p}{\to}\pi^+{n}$ data as explained in the text.}
\label{gpion23}
\end{figure}

The differential cross sections for $\gamma{n}{\to}\pi^-p$ 
are presented in Figs.~\ref{gpion22} and \ref{gpion23} as a function of $-t$ 
for different photon energies. 
The measurements at JLab were done at 
$E_\gamma{\simeq}$1.1, 1.65, 1.8, 2.48, 3.3,
4.1 and 5.53 GeV, which correspond to 
$\sqrt{s}{\simeq}$1.7, 1.99, 2.06, 2.35, 2.67, 2.95 
and 3.36~GeV, respectively. For completeness and for 
illustrating the compatibility
with other available experimental results we also show 
differential cross sections from 
Refs.~\cite{Heide,BarYam,Scheffler,Sternemann} obtained 
at almost the same photon energies.

It is instructive to recall here the results of our 
analysis of the $\gamma{p}{\to}\pi^+n$ 
data by the Hall A Collaboration~\cite{Zhu,Zhu1} 
at the same photon energies and the same 
range of $t$, shown in Figs.~\ref{gpion21} and \ref{gpion24}. 
There, we found that at $\sqrt{s}$=1.7~GeV our 
calculation substantially underestimates 
the $\pi^+$ spectrum and we observe a similar deficiency now for $\pi^-$-meson 
photoproduction. This discrepancy is most likely 
associated with contributions from known 
resonances in that energy region which are missing in our model calculation. 
At the energies $\sqrt{s}$=1.99 and 2.06 GeV the positive 
photoproduction spectra at $|t|{\le}0.7$~GeV$^2$ is reasonably described
by our model calculation, and for the energies 2.35 and 2.67~GeV even up 
to roughly $|t|{=}1.5$~GeV$^2$. (There are no experimental points at 
$|t|{\le}$2~GeV$^2$ for $\sqrt{s}$=2.95 and 3.36~GeV.) Interestingly, 
the situation for negative pion photoproduction 
is somewhat different. While the 
model reproduces the $\gamma{n}{\to}\pi^-p$ differential cross sections at 
$\sqrt{s}{\simeq}$2.67 and 2.95 GeV quite well up to $|t|{\approx}1.5$~GeV$^2$, 
we observe a much more substantial deviation at the lower energies and for 
$|t|{\ge}0.7$~GeV$^2$.
In particular, at $\sqrt{s}$=1.99, 2.06 and 2.35~GeV 
the $t$ dependence of negative pion photoproduction 
differs drastically from that for 
positive pions for $|t|{\ge}0.5$~GeV$^2$, say. In fact, within the range 
$0.5{<}|t|{<}2$~GeV$^2$ where the JLab data are available, the differential 
cross sections for $\gamma{n} \to \pi^-p$ are practically independent of the 
four-momentum transfer squared.
Data from other experiments~\cite{Sternemann,Benz} 
exhibit a comparable behavior 
although they are afflicted by large uncertainties. Note that a very similar 
$t$ dependence was observed in negative pion photoproduction~\cite{BarYam}
at $E_\gamma$=3~GeV, or $\sqrt{s}$=2.55~GeV, presented in Fig.~\ref{gpion14}.
This could be an indication for contributions from excited 
baryons with masses lying around 1.99${\le}\sqrt{s}{\le }$2.55 GeV. The range 
seems to be too large for a single resonance, 
unless one assumes the contribution 
to be from a rather broad ($\simeq$600~MeV) structure. 

We note that the GWU PWA~\cite{Said3,Said1,Said2} 
reproduces nicely $\pi^+$ as well as 
$\pi^-$-meson photoproduction data at $\sqrt{s}{\le}$2.1~GeV. 
In particular, it describes the flat
$t$-dependence for negative pions. 
It is unlikely that $U$ symmetry was implemented in this
analysis and the most natural expectation is that the PWA of the data yields a 
much larger photon coupling to the neutron than to the 
proton for resonances located within the
range of 1.99${\le}\sqrt{s}{\le }$2.55~GeV. 
Indeed the SM-95 solution~\cite{Said3}
finds evidence for the excited baryons $F_{35}(1905)$, $D_{35}(1930)$ 
and $F_{37}$ $(1950)$, 
but the results for the $\gamma{n}$ couplings are not 
given in the corresponding publication.
In this context, let us mention that it was shown 
within the $1/N_c$ expansion, based on the approximate dynamical spin-flavour
symmetry $SU(4)$ of QCD in the large $N_c$ limit~\cite{Goity1,Goity2},  
that the photoproduction on the neutron can be larger than that on the proton. 
Thus, it is conceivable that the chances for 
exciting a baryon in $\gamma{n}{\to}\pi^-p$ 
are substantially larger than in 
the $\gamma{p}{\to}\pi^+n$ or $\gamma{p}{\to}\pi^0p$ 
reactions. Furthermore, according to 
the systematic study of Ref.~\cite{Goity1} one should expect 
that such an excited baryon 
is a nucleon, because photo-excitation of $\Delta$ 
resonances should be identical for 
proton and neutron targets.
	     
The presently available data are too scarce to allow us 
to draw a definitive conclusion. 
Apparently new precise measurements at $|t| < $ 2~GeV$^2$ and photon energies 
1.6${\le}E_\gamma{\le}$3.4~GeV are required to clarify the situation. 
At such energies this $t$ range is quite promising for baryon spectroscopy,
because at large $-t$ the contribution from hard QCD processes might dominate 
the reaction.

\section{The {\boldmath$\pi^-{/}\pi^+$} ratio}

Quite interesting information on charged pion 
photoproduction is provided by the 
ratio ${\cal R}$ of the $\gamma{n}{\to}\pi^-{p}$ 
to $\gamma{p}{\to}\pi^+{n}$ differential 
cross section as a function of $t$ and the photon energy or $\sqrt{s}$. 
Since at small $|t|$ ($|t|{\le}m_\pi^2$) single pion 
photoproduction is dominated 
by $t$-channel pion exchange, it follows that ${\cal R}{=}1$ -- independent of 
the energy.  At moderate $t$ the interference between the 
$\pi$ and $\rho$ exchanges is expected to result in a decrease of ${\cal R}$ 
as $|t|$ increases, following Eq.~(\ref{gpar}). With further increase of $|t|$
the contribution of pion exchange vanishes and 
$\rho$ exchange dominates so that
one might expect a return to ${\cal R}{=}1$. However, since other contributions,
summarized in Table~\ref{traj}, could be sizeable the evolution of ${\cal R}$ 
with $t$ is not trivial. Thus, this evolution directly reflects the presence of
contributions to the reaction amplitude from exchanges 
with different quantum numbers. 

Note that the $\pi^-{/}\pi^+$ ratio at large $|t|$ can 
be compared with the handbag 
calculations~\cite{Afanasev,Huang1,Huang2} based 
on hard gluon exchange. Therefore, it 
is important to inspect the behavior of ${\cal R}$ 
when approaching $-t{\simeq}$2~GeV$^2$. 
Here one expects the transition between perturbative QCD as modelled by Regge theory 
and hard QCD processes.

\begin{figure}[b]
\vspace*{-6.5mm}
\centerline{\hspace*{6mm}\psfig{file=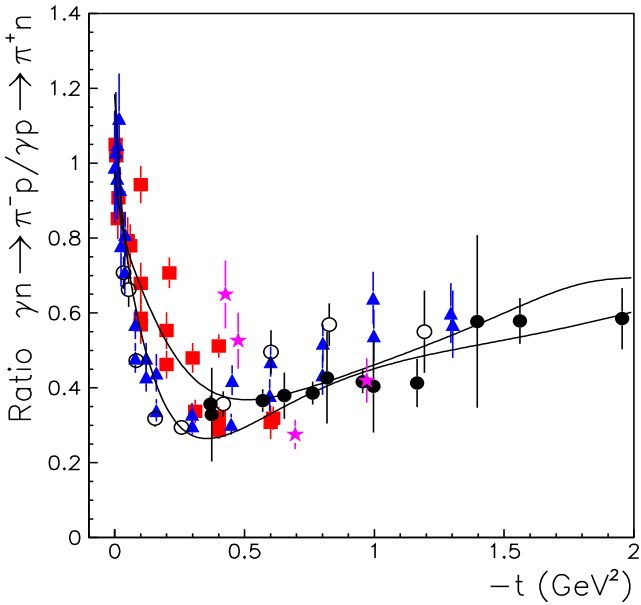,width=9.6cm,height=9.cm}}
\vspace*{-3mm}
\caption{The ratio of the $\gamma{n}{\to}\pi^-{p}$ to $\gamma{p}{\to}\pi^+{n}$
differential cross section as a function of $-t$. 
The data for 3.4${\le}E_\gamma{\le}$16~GeV are taken from 
Refs. \cite{Heide} (filled squares), \cite{Sherden} (open circles),
\cite{BarYam} (filled circles) and \cite{Boyarski2} (filled triangles). 
The two solid lines show  our results obtained for $E_\gamma$=3.4 and 16~GeV.
}
\label{gpion1}
\end{figure}

{}Figure~\ref{gpion1} shows the ratio ${\cal R}$ of 
the $\gamma{n}{\to}\pi^-{p}$ to 
$\gamma{p}{\to}\pi^+{n}$ differential cross section as a function of the 
four-momentum transfer squared. Here we include data for photon energies 
3.4$\le E_\gamma \le $16~GeV. In each of the 
experiments~\cite{Sherden,Heide,BarYam,Boyarski2} the ratio ${\cal R}$ was
measured for 
a fixed photon energy as a function of $t$ 
or of the pion production angle, $\theta^\ast$. 
The data exhibit a very specific $t$-dependence, 
that is almost independent of the 
energy. 
Approaching $t=0$ the ratio ${\cal R}$ is close to 1, as expected from the 
dominance of pion exchange at $|t|{\le}m_\pi^2$. 
Then the ratio decreases because of the 
interplay between the various contributions to the 
photoproduction amplitude listed in Table \ref{traj} 
and entering Eq.~(\ref{gpar}).
However, with increasing $|t|$ the ratio does 
not converge to unity as one might 
expect from the dominance of $\rho$ exchange. This clearly indicates that
with increasing $-t$ the reaction is still governed by contributions from 
several different processes and that one will not able to reproduce such a 
$t$ dependence within a simple $\pi{+}\rho$ model.

The $\pi^-{/}\pi^+$ ratio was also measured recently at JLab by the 
Hall A Collaboration~\cite{Zhu,Zhu1}. As mentioned above, 
this experiment was motivated by hard QCD physics~\cite{Brodsky,Matveev} 
and devoted to pion photoproduction at large $|t|$. Although it is difficult to 
provide an estimate for the absolute value of the reaction cross section 
within QCD inspired models, predictions for the $\pi^-{/}\pi^+$ 
ratio and for some polarization observables at large $|t|$ can 
be made with more  
confidence~\cite{Huang1,Huang2,Afanasev}. Indeed, the calculations
of Refs.~\cite{Huang1,Huang2} reproduce ${\cal R}$ at large $-t$ rather well.
Part of the data were also taken for $|t|{<}$2~GeV$^2$, 
which allows us to compare  
those data with our calculation and to search for a 
signature~\cite{Brodsky,Matveev,Huang1,Huang2} of the transition from pQCD,  
modelled by Regge theory, to hard QCD. 

On the other hand, 
the JLab experiment~\cite{Zhu,Zhu1} was done at different 
photon energies and for fixed 
angles $\theta^\ast$ in the overall cm system, 
which complicates the comparison with  
other results. Specifically, it is not possible to 
evaluate the $t$ dependence of
the $\pi^-{/}\pi^+$ ratio from these data and compare 
it with either that of the other data sets or  
of our model. 

\begin{figure}[t]
\vspace*{-6.mm}
\centerline{\hspace*{6mm}\psfig{file=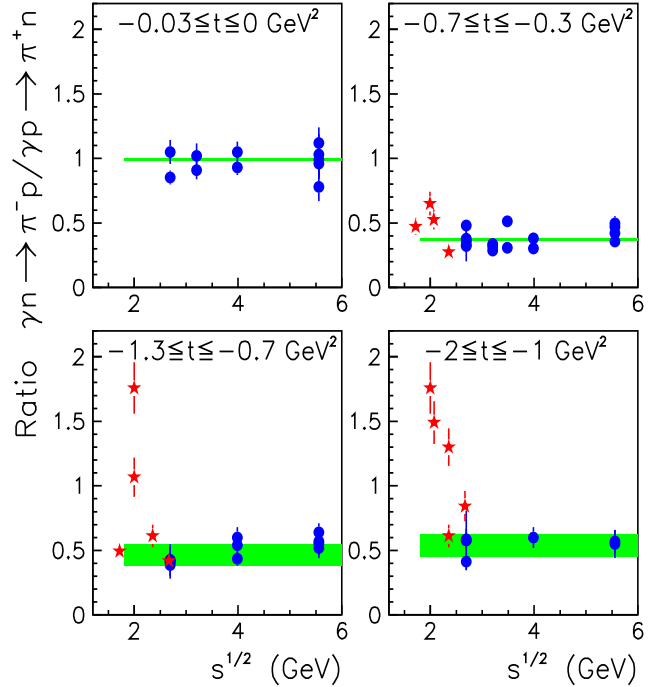,width=9.6cm,height=10.5cm}}
\vspace*{-5mm}
\caption{The ratio of the $\gamma{n}{\to}\pi^-{p}$ to $\gamma{p}{\to}\pi^+{n}$
differential cross section as a function of invariant collision energy shown for
different intervals of the four-momentum transfer squared. 
The filled circles are experimental results from  
Refs.~\cite{Heide,Sherden,BarYam,Boyarski2}, while the stars indicate
data from JLab~\cite{Zhu,Zhu1}. The bands show the variation
of ${\cal R}$ within the indicated range of $t$ as predicted by our model.}
\label{gpion1a}
\end{figure}

The solid lines in Fig.~\ref{gpion1} show our results for $E_\gamma$=3.4 
and 16~GeV. The model reproduces the $t$ dependence qualitatively and 
exhibits only a mild dependence on energy. 
Note that within this energy range the differential 
cross section itself changes by almost two orders of 
magnitude, as is visible in 
Figs.~\ref{gpion3} and \ref{gpion15_r1}. 

In any case, we can directly test the model 
by considering the $\sqrt{s}$ dependence 
of the ratio ${\cal R}$ at fixed values of $t$. 
Since the data are not available at exactly 
the same $t$ one can select appropriate ranges. 
This is done in Fig.~\ref{gpion1a},  
where we display the dependence of ${\cal R}$ on the invariant collision 
energy $\sqrt{s}$ for different intervals of $t$. The band indicates the 
variation of ${\cal R}$ as predicted by the 
model for the selected range of $t$. 
 
Obviously the $\pi^-{/}\pi^+$ ratio obtained from experimental results available
at -0.03${\le}t{\le}$0~GeV$^2$ is close to unity at energies 
2.7${\le}\sqrt{s}{\le}5.6$~GeV. That is exactly what one would expect from 
the pion exchange dominance at $|t|{\le}m_\pi^2$. This feature is reproduced by
the model.
With regard to other intervals of $t$ which we have considered,  
the data above $\sqrt{s}{\simeq}$2.5~GeV from 
Refs.~\cite{Sherden,Heide,BarYam,Boyarski2} are well described by our
model. Furthermore, the JLab data~\cite{Zhu,Zhu1} available at the same 
energies are in good agreement with other data and also with our calculation.
However, in the range 1.7${\le}\sqrt{s}{\le}2.5$ GeV, the ratio of the 
$\gamma{n}{\to}\pi^-{p}$ to $\gamma{p}{\to}\pi^+{n}$ differential 
cross section shows a clear resonance-like structure, which is most prominently
noticeable at 0.7${\le}|t|{\le}$2 GeV$^2$. This observation is consistent with 
the conclusions we drew from our analysis of 
the $\pi^-$ differential cross section
above. 

\section{Conclusion}
We analyzed the data on charged pion photoproduction 
available at photon energies $3{\le}E_\gamma{\le}8$~GeV and at four-momentum 
transfer squared $|t|{\le}$2~GeV$^2$ within the Regge approach.
The model was constructed by taking into account both pole and cut exchange 
$t$-channel helicity amplitudes. 
We consider the $b_1$, $\rho$ and $a_2$ trajectories 
and pion exchange and fix the unknown model parameters such 
as the helicity couplings by fitting experimental results on differential cross 
sections, the polarized photon asymmetry and recoil and target asymmetries.

The model provides a reasonable description of the data, indicating that for 
the energy range considered single pion photoproduction is dominated by 
nonresonant contributions. The calculation was extended to 
lower photon energies in order to examine the data with regard to possible 
signals for the excitation of baryonic resonances 
with masses between 2 and 3~GeV. 
We detected a systematic discrepancy between the calculation and the data 
on $\gamma{n}{\to}\pi^-{p}$ differential 
cross sections for photon energies from 
1.65 to 3~GeV (invariant collision 
energies of 1.99$ \le \sqrt{s} \le 2.55$~GeV) 
in the region $-t{\ge}$0.5 GeV$^2$. 
The model results for $\gamma{p}{\to}\pi^+{n}$ also show deviations from 
the data in this energy and $t$ region, though here the disagreement is less
pronounced.  

The differential cross sections for $\gamma{n}{\to}\pi^-{p}$ 
which are at variance 
with the model calculation are those measured at 
ELSA (Bonn)~\cite{Sternemann} and very recently at JLab~\cite{Zhu1}. 
Unfortunately, the amount and accuracy of the experimental results in the
relevant energy region is still insufficient for a more detailed quantitative 
analysis and for the evaluation of possible 
contributions from the excitation of 
high-mass baryons. Nevertheless, we observe a resonance-like structure in 
the ratio of the $\gamma{n}{\to}\pi^-{p}$ to 
$\gamma{p}{\to}\pi^+{n}$ differential cross sections taken at fixed intervals of
$t$ and shown as a function of $\sqrt{s}$. 
This ratio exhibits a noticeable enhancement
at 1.7${\le}\sqrt{s}{\le}2.5$~GeV as compared to lower and higher energies.

Our findings suggest that the prospects for 
the excitation of baryon resonances on neutrons 
via photons could be substantially larger than on protons. 
If this is the case, it 
will be more difficult to observe such resonance excitations in the 
$\gamma{p}{\to}\pi^0p$ reaction.
Evidently, the validity of this conjecture can 
be examined via the analysis of data 
reported very recently~\cite{Pee} by the CB-Collaboration at ELSA. 
Note that in the framework of the $1/N_c$ expansion based on the 
approximate dynamical spin-flavour symmetry, $SU(4)$, 
of QCD in the large $N_c$ limit, it 
was shown~\cite{Goity1,Goity2} that photoproduction on 
the neutron can be very different 
from that on the proton. Furthermore, according to the systematic study of 
Ref.~\cite{Goity1} one might expect that such an excited baryon is a nucleon,
because photo-excitation of $\Delta$ resonances is identical for proton and 
neutron targets.

Further progress in understanding the observed 
discrepancies requires new dedicated 
experiments on the $\gamma{n}{\to}\pi^-p$ and $\gamma{n}{\to}\pi^0n$ reactions 
at photon energies 1.6${\le}E_\gamma{\le}$3~GeV. 
Apparently polarization measurements
are necessary to enable a reconstruction of the quantum numbers 
of the excited baryons.

\subsection*{Acknowledgements}
This work was partially  supported  by Deutsche
Forschungsgemeinschaft  through funds provided to the SFB/TR 16
``Subnuclear Structure of Matter''. This research is part of the \, EU
Integrated \, Infrastructure \, Initiative Hadron Physics Project under
contract  number RII3-CT-2004-506078. This work was also supported in 
part by U.S. DOE Contract No. DE-AC05-06OR23177, under which 
Jefferson Science Associates, LLC, operates Jefferson Lab.
A.S. acknowledges support by the
JLab grant SURA-06-C0452  and the 
COSY FFE grant No. 41760632 (COSY-085).


\begin{thebibliography}{99}
\bibitem{Isgur}
	N. Isgur and G. Karl, Phys. Rev. D {\bf 18}, 4187 (1978) .
\bibitem{Capstick}
	S. Capstick and N. Isgur, Phys. Rev. D {\bf 34}, 2809 (1986).	
\bibitem{Capstick1}
	S. Capstick and W. Roberts, Phys. Rev. D {\bf 47}, 1994 (1993).
\bibitem{Capstick2}
	S. Capstick and W. Roberts, Phys. Rev. D {\bf 49}, 4570 (1994)
        [arXiv:nucl-th/9310030] . 
\bibitem{Shifman}
        M.A. Shifman, A.I. Vainshtein and V.I. Zakharov,
	Nucl. Phys. B {\bf 147},  385 (1979).
\bibitem{Ioffe1}
	B.L. Ioffe, Nucl. Phys. B {\bf 188},  317 (1981).
\bibitem{Meissner1}
	Ulf-G. Mei{\ss}ner, {\it Encyclopedia of Analytic QCD}, ed. 
	M. Shifman, World Scientific [arXiv:hep-ph/0007092].
\bibitem{Ioffe2}
	B.L. Ioffe, Prog. Part. Nucl. Phys. {\bf 56},  232 (2006)
	[arXiv:hep-ph/0502148].
\bibitem{GellMann}
	M. Gell-Mann and M. Levy, Nuovo Cim. {\bf 16}, 705  (1960).
\bibitem{Nambu}
	Y. Nambu and G. Jona-Lasino, Phys. Rev. {\bf 122}, 345 (1961).
\bibitem{Barger1}
	V. Barger and D. Cline, Phys. Rev. Lett. {\bf 16}, 913 (1966).
\bibitem{Barger2}
	V. Barger and M. Olsson, Phys. Rev. {\bf 151}, 1123 (1966).
\bibitem{Collins4}
	P.D.B. Collins, {\it Regge Theory and High Energy Physics}, 
	Cambridge, Cambridge University Press (1977).
\bibitem{Iwasaki}
	Y. Iwasaki, Prog. Theor. Phys. {\bf 44},  777 (1970) .
\bibitem{Iachello}
	F. Iachello, Phys. Rev. Lett. {\bf 63},   1891 (1989).
\bibitem{Robson}
	D. Robson, Phys. Rev. Lett. {\bf 63}, 1890 (1989).
\bibitem{Kirchbach1}
	M. Kirchbach, Mod. Phys. Lett. A {\bf 12}, 3177 (1997)
	[arXiv:nucl-th/9712072].
\bibitem{Kirchbach2}
	M. Kirchbach, Int. J. Mod. Phys. A {\bf 15},  1435 (2000)
	[arXiv:nucl-th/0007022].
\bibitem{Collins:1976yq}
  J.~C.~Collins, A.~Duncan and S.~D.~Joglekar,
  Phys.\ Rev.\  D {\bf 16}, 438 (1977).
\bibitem{Ji:1994av}
  X.~D.~Ji,
  Phys.\ Rev.\ Lett.\  {\bf 74}, 1071 (1995)
  [arXiv:hep-ph/9410274].
\bibitem{Yao}	
	W.-M. Yao {\it et al.}, J. Phys. G {\bf 33}, 1  (2006).
\bibitem{Glozman1}
  L.Y.~Glozman, Phys.\ Rept.\  {\bf 444}, 1 (2007)
  [arXiv:hep-ph/0701081].
\bibitem{Glozman2}
	L.Y. Glozman, Phys. Lett. B {\bf 475},  329 (2000) [arXiv:hep-ph/9908207].
\bibitem{Jido1}
	D. Jido, T. Hatsuda  and T. Kunihiro, Phys. Rev. Lett. {\bf 84}, 
	3252 (2000) [arXiv:hep-ph/9910375].
\bibitem{Jido2}
	D. Jido, M. Oka and A.Hosaka, Prog. Theor. Phys. {\bf 106},
	873 (2001) [arXiv:hep-ph/0110005].
\bibitem{Cohen1}
	T.D. Cohen and L.Y. Glozman, Phys. Rev. D {\bf 65}, 016006 (2002)
	[arXiv:hep-ph/0102206].
\bibitem{Cohen2}
	T.D. Cohen and L.Y. Glozman, Int. J. Mod. Phys. A  {\bf 17},
	1327 (2002) [arXiv:hep-ph/0201242].
\bibitem{Cohen3}
	T.D. Cohen, Nucl. Phys. A {\bf 775}, 89 (2006) [arXiv:hep-ph/0605206].
\bibitem{Jaffe1}
	R.L. Jaffe, D. Pirjol and  A. Scardicchio, Phys. Rev. Lett. {\bf 96}
	 121601 (2006) [arXiv:hep-ph/0511081].
\bibitem{Jaffe2}	
	R.L. Jaffe, D. Pirjol and A. Scardicchio, Phys. Rept. {\bf 435 }, 
	157 (2006)  [arXiv:hep-ph/0602010].	
\bibitem{Gonzalez}
	P. Gonzalez, J. Vijande, A. Valcarce, and H. Garcilazo, 
 Eur.\ Phys.\ J.\  A {\bf 31}, 515 (2007)
  [arXiv:hep-ph/0610257].
\bibitem{Loring:2001kx}
  U.~L\"oring, B.~C.~Metsch and H.~R.~Petry,
  Eur.\ Phys.\ J.\  A {\bf 10}, 395 (2001) 
  [arXiv:hep-ph/0103289].

\bibitem{Loring:2001ky}
  U.~L\"oring, B.~C.~Metsch and H.~R.~Petry,
  Eur.\ Phys.\ J.\  A {\bf 10}, 447 (2001)
  [arXiv:hep-ph/0103290].

\bibitem{Loring:2001bp}
  U.~L\"oring and B.~Metsch,
  arXiv:hep-ph/0110412.
\bibitem{Hoehler1}
	G. H{\"o}hler, $\pi{N}$ Newsletter {\bf 9}, 1 (1993).
\bibitem{Hoehler2}
	G. H{\"o}hler, Landolt-B{\"o}rnstein {\bf 9}, Springer, Berlin, 1983.
\bibitem{Koch}
	R. Koch, Nucl. Phys. A {\bf 448}, 707 (1986).
\bibitem{Cutkovsky}
	R. E. Cutkovsky {\it et al.},  Phys. Rev. D {\bf 20}, 2839 (1980).
\bibitem{Hendry}
	A.W. Hendry, Phys. Rev. Lett. {\bf 41}, 222 (1978).
\bibitem{Manley}
	D. M. Manley and E. M. Saleski, Phys. Rev. D {\bf 45}, 4002 (1992).
\bibitem{Arndt96}
	R.A. Arndt, W.J. Briscoe, I.I. Strakovsky and R.L. Workman. Phys. Rev.
	C {\bf 74}, 045205 (2006) [arXiv:nucl-th/0605082].
\bibitem{GWU}
	SAID Partial Wave Analyis; the current solution is available
	on the web: http://gwdac.phys.gwu.edu/.
\bibitem{leesmith-07}
	T.-S. H. Lee and L. C. Smith, 
        J.\ Phys.\ G {\bf 34}, S83 (2007) [arXiv:nucl-th/0611034].
\bibitem{julich1}
  O.~Krehl, C.~Hanhart, S.~Krewald and J.~Speth,
  Phys.\ Rev.\  C {\bf 62}, 025207 (2000)
  [arXiv:nucl-th/9911080].
\bibitem{julich2}
  A.~M.~Gasparyan, J.~Haidenbauer, C.~Hanhart and J.~Speth,
  Phys.\ Rev.\  C {\bf 68}, 045207 (2003)
  [arXiv:nucl-th/0307072].
\bibitem{Yang2006}
  S.~N.~Yang, G.~Y.~Chen and S.~S.~Kamalov,
  Nucl.\ Phys.\  A {\bf 790}, 229 (2007)
  [arXiv:nucl-th/0610076].
\bibitem{Chen2007}
  G.~Y.~Chen, S.~S.~Kamalov, S.~N.~Yang, D.~Drechsel and L.~Tiator,
  arXiv:nucl-th/0703096.
\bibitem{Julia2007}
 B. Juli\'a-Diaz, T.-S. H. Lee, A. Matsuyama, and T. Sato, 
 arXiv:0704.1615 [nucl-th].
\bibitem{Gross2}
	     Y. Surya and  F. Gross,  Phys. Rev. C {\bf 53},  2422 (1996).
\bibitem{Sato}
	     T. Sato and  T.S.H. Lee, Phys. Rev. C {\bf 54},
	     2660 (1996) [arXiv:nucl-th/9606009].
\bibitem{Fuda}
	     M.G. Fuda and H. Alharbi, Phys. Rev. C {\bf 68},   064002 (2003).
\bibitem{Pascalutsa}
  V.~Pascalutsa and J.~A.~Tjon,
  Phys.\ Rev.\  C {\bf 70}, 035209 (2004)
  [arXiv:nucl-th/0407068].
\bibitem{Haberzettl}
  H.~Haberzettl, K.~Nakayama and S.~Krewald,
  Phys.\ Rev.\  C {\bf 74}, 045202 (2006)
  [arXiv:nucl-th/0605059].
\bibitem{msl}
     A. Matsuyama, T. Sato, T.- S. H. Lee, Phys. Rept.
{\bf 439}, 193 (2007) [arXiv:nucl-th/0608051].
\bibitem{Feuster1}
    T. Feuster and U. Mosel, Phys. Rev. C {\bf 59}, 460 (1999)
    [arXiv:nucl-th/9803057].
\bibitem{Penner}
	G. Penner and U. Mosel, Phys. Rev. C {\bf 66},
	055211 (2002) [arXiv:nucl-th/0207066].
\bibitem{Klempt}
  A.~Anisovich, E.~Klempt, A.~Sarantsev and U.~Thoma,
  Eur.\ Phys.\ J.\  A {\bf 24}, 111 (2005)
  [arXiv:hep-ph/0407211].
\bibitem{Klempt1}
  A.~V.~Anisovich, A.~Sarantsev, O.~Bartholomy, E.~Klempt, V.~A.~Nikonov and U.~Thoma,
  Eur.\ Phys.\ J.\  A {\bf 25}, 427 (2005)
  [arXiv:hep-ex/0506010].
\bibitem{Perl}
	M.L. Perl, {\it High Energy Hadron Physics}, New York, Wiley (1974) 395.
\bibitem{Collins2}
	P.D.B. Collins, {\it An Introduction to Regge Theory and High Energy
	Physics}, Cambridge University, Cambridge, England (1977) 275.
\bibitem{Collins3}
	P.D.B. Collins and  A.D. Martin, Rept. Prog. Phys. {\bf 45},
	335 (1982).
\bibitem{Caneschi}
	     L.  Caneschi, {\it Regge Theory of Low $p_t$ Hadronic
	     Interactions} Amsterdam, North-Holland (1989)	    
\bibitem{Matthiae}
	     G. Matthiae, Rep. Prog. Phys. {\bf 57},  743 (1994).
\bibitem{Levin}
        E. Levin, arXiv:hep-ph/9710546.
\bibitem{Zhu}
	L. Y. Zhu {\it et al.}, Phys. Rev. C {\bf  71}, 
	044603 (2005) [arXiv:nucl-ex/0409018].
\bibitem{Zhu1}
	L. Y. Zhu {\it et al.}, Phys. Rev. Lett. {\bf 91}, 
	022003 (2003) [arXiv:nucl-ex/0211009]
\bibitem{Rahnama2}
	M. Rahnama and J. K. Storrow, J. Phys. G {\bf 17},  243 (1991).
\bibitem{Storrow1}	
	J. Storrow, {\it Electromagnetic Interactions of Hadrons}, ed. A.
Donnache and G. Shaw, {\bf 1}, New York Plenum (1978).
\bibitem{Kellett}
	B. H. Kellett, Nucl. Phys. B {\bf 25}, 205 (1970).
\bibitem{Ball1}
	J.S. Ball, W.R. Frazer and M. Jacob, Phys. Rev. Lett. {\bf 20},
	518 (1968).	
\bibitem{Henyey1}	
	F. Henyey, 
	Phys. Rev. {\bf 170},   1619 (1968).
\bibitem{Ball}
	J. S. Ball, H. W. M\"uller and B. K. Pal, Phys. Rev. D {\bf  4}, 
	2065 (1971).
\bibitem{Jackson}
	J. D. Jackson and C. Quigg, Phys. Lett. B {\bf  29},   236 (1969).
\bibitem{Jackson1}
	J. D. Jackson and C. Quigg, Phys. Lett. B {\bf  33},  444 (1970).
\bibitem{Froyland}
	J. Froyland and D. Gordon, Phys. Rev. {\bf 177},  2500 (1969).
\bibitem{Rahnama1}
	M. Rahnama and J. K. Storrow, J. Phys. G. {\bf 8}, 455 (1982).
\bibitem{Guidal}
	M. Guidal, J.M. Laget and  M. Vanderhaeghen,
	Nucl. Phys. A {\bf 627},  645 (1997).
\bibitem{Vanderhaeghen}
	M. Vanderhaeghen, M. Guidal and  J.M. Laget,
	Phys. Rev. C {\bf 57}, 1454 (1998).	
\bibitem{Dombey1}
	N. Dombey, Nuovo Cim. {\bf 32},   1696 (1964).
\bibitem{Dombey2}
	N. Dombey, Phys. Lett. B {\bf 30}, 646 (1964).	
\bibitem{Kellett1}
	B. H. Kellett, Nucl. Phys. B {\bf  35},  517 (1971).
\bibitem{Blackmon}
	M. L. Blackmon, G. Kramer and K. Schilling, Nucl. Phys. B {\bf  12},
	495 .(1969) 
\bibitem{Kramer1}
	G. Kramer and P. Stichel, Z. Phys. {\bf 178},  519 (1964).
\bibitem{Berends1}
	F.A. Berends, Phys. Rev. D {\bf 1},  2590 (1970).
\bibitem{Berends2}
	F.A. Berends and  G.B. West, Phys. Rev. {\bf 188}, 2538  (1969).
\bibitem{Dolen}
	R. Dolen, D. Horn and C. Schmid, Phys. Rev. {\bf 166},  1768 (1968).
\bibitem{Said3}
	 R. A. Arndt, I. I. Strakovsky and R. L. Workman, Phys. Rev. 
	{\bf C 53} (1996) 430; updates available on the web:
	http://gwdac.phys.gwu.edu/
\bibitem{Irving}
	A. C. Irving and R. P. Worden, Phys. Rept. {\bf 34}, 117 (1977).	
\bibitem{Sopkovitch}
	N.J. Sopkovitch, Nuovo Cim. {\bf 26},   186 (1962).
\bibitem{Gottfried}
	K. Gottfried and J.D. Jackson, Nuovo Cim. {\bf 34},   735 (1964).
\bibitem{Jackson3}
	J.D. Jackson, Rev. Mod. Phys. {\bf 42},  12 (1970).
\bibitem{Worden}
	R. Worden, Nucl. Phys. B {\bf 37},   253 (1972).
\bibitem{Arnold}
	R.C. Arnold, Phys. Rev. {\bf 153},  1523 (1967).
\bibitem{Gribov}
	V.M. Gribov, I.Y. Pomeranchuk and K.A. Ter-Martirosyan, 
	Phys. Rev. {\bf 139B},  184 (1965).
\bibitem{White3}
	A.R. White, Nucl. Phys. B {\bf 50},   93 (1972). 
\bibitem{White4}
	A.R. White, Nucl. Phys. B {\bf 50},  130 (1972). 
\bibitem{Chew}
	G. F. Chew, M. L. Goldberger, F. E. Low and Y. Nambu, Phys. Rev. {\bf
	106}, 1345  (1957).
\bibitem{Wiik}	
	B.H. Wiik,  Proc. Int. Symp. on Electron and Photon
	Interactions at High Energies, Ithaca, New York, Aug 1971, 163.
\bibitem{Baker}
	I. S  Barker, A. Donnachie and J. K. Storrow, Nucl. Phys. B {\bf 95},
	347 (1975).
\bibitem{Berends}		
	F. A. Berends, A. Donnachie and D. L. Weaver, Nucl. Phys. B {\bf 4},
	1 (1967).
\bibitem{Said1}
	R. A. Arndt, R. L. Workman, Z. Li and L. D. Roper, Phys. Rev. 
	C {\bf  42}, 1853 (1990) .
\bibitem{Chiu}
	C.B. Chiu and S. Matsuda, Phys. Lett. B {\bf 31},  455 (1970).
\bibitem{Sibirtsev2}
	A. Sibirtsev, K. Tsushima and  S. Krewald, Phys. Rev.
	C {\bf 67}, 055201 (2003) [arXiv:nucl-th/0301015].
\bibitem{White1}
	J. N. J. White, Phys. Lett. B {\bf  26}, 461 (1968).	
\bibitem{White2}
	J. N. J. White, Nucl. Phys. B {\bf 13},  139  (1969).
\bibitem{Henyey}	
	F. Henyey, G. L. Kane, Jon Pumplin and M. H. Ross,
	Phys. Rev. {\bf 182}, 1579 (1969).
\bibitem{Kramer2}
	G. Kramer, K. Shilling and L. Stodolsky, Nucl. Phys. B {\bf 5}, 
	317 (1968).
\bibitem{Dombey}
	     N. Dombey, Phys. Lett. B {\bf 30},  646 (1969).
\bibitem{Durham}
         The Durham High Energy Physics  Databases; the data are available on 
         the web: http: //durpdg.dur.ac.uk/HEPDATA/.
\bibitem{Rahnama3}
	M. Rahnama and J. K. Storrow, Z. Phys. C {\bf  10},  263 (1981).
\bibitem{Baker1}
	I. S  Barker, A. Donnachie and J. K. Storrow, Nucl. Phys. B {\bf 79},
	 431 (1974).
\bibitem{Ecklund}
	S. D. Ecklund and R. L. Walker, Phys. Rev. {\bf  159},  1195 (1967). 
\bibitem{Buschhorn1}
	G. Buschhorn {\it et al.}, Phys. Rev. Lett. {\bf 17}, 1027 (1966).
\bibitem{Buschhorn2}
	G. Buschhorn {\it et al.}, Phys. Rev. Lett. {\bf 18}, 571 (1967).
\bibitem{Dowd}
	J. P. Dowd, D. O. Caldwell, K. Heinloth and T. R. Sherwood,
	Phys.\ Rev. \ Lett.\ {\bf 18}, 414 (1967) .
\bibitem{Heide}
	P. Heide, U. K\"otz, R. A. Lewis, P. Schm\"user, H. J. Skronn and
	H.~Wahl, Phys. Rev. Lett. {\bf 21}, 248  (1968).
\bibitem{Joseph}
	P. M. Joseph, N. Hicks, L. Litt, F. M. Pipkin and J. J. Russell,
	Phys.\ Rev. \ Lett.\ {\bf 19}, 1206  (1967).
\bibitem{BarYam}
	Z. Bar-Yam {\it et al.}, Phys. Rev. Lett. {\bf 19},  40 (1967).
\bibitem{Elings}
	V. B. Elings {\it et al.}, Phys. Rev. {\bf  156},  1433 (1967).
\bibitem{Anderson1}
	R. L. Anderson {\it et al.}, Phys. Rev. D {\bf  14}, 679 (1976).
\bibitem{Anderson2}
	R. L. Anderson {\it et al.}, Phys. Rev. Lett. {\bf 23},  721 (1969).
\bibitem{Boyarski1}
	A. M. Boyarski {\it et al.}, Phys. Rev. Lett. {\bf 20},  300  (1968).
\bibitem{Geweniger}
	C. Geweniger {\it et al.}, Phys. Lett B {\bf  29},  41 (1969).	
\bibitem{BarYam2}
	Z. Bar-Yam {\it et al.}, Phys. Rev. Lett. {\bf 25},   1053 (1970).
\bibitem{Sherden}
	D. J. Sherden {\it et al.}, Phys. Rev. Lett. {\bf 30},  1230 (1973).
\bibitem{Bussey1}	
	P.J. Bussey {\it et al.}, Nucl. Phys. B {\bf  154}, 492 (1979).
\bibitem{Bienlein}
	H. Bienlein {\it et al.}, Phys. Lett. B {\bf  46},  131 (1973).
\bibitem{Booth}	
	P.S.L. Booth {\it et al.}, Phys. Lett. B {\bf  38}, 339 (1972).
\bibitem{Deutsch}
	M. Deutsch, L. Golub, P. Kijewski, D. Potter, D.J. Quinn and 
	J. Rutherfoord, Phys. Rev. Lett. {\bf 29}, 1752 (1972).
\bibitem{Genzel}
	H. Genzel {\it et al.}, Nucl. Phys. B {\bf   92}, 196 (1975).	
\bibitem{Morehouse}	
	C. C. Morehouse	{\it et al.}, Phys. Rev. Lett. {\bf 25}, 835 (1970) .
\bibitem{Bussey2}
	P.J. Bussey {\it et al.}, Nucl. Phys. B {\bf  154},   205 (1979).	
\bibitem{Burfeindt}	
	H. Burfeindt {\it et al.}, Phys. Lett. B {\bf  33},   509 (1970).
\bibitem{Stichel}	
	P. Stichel, Z. Phys. {\bf 180}, 170 (1964).	
\bibitem{Ravndal}
	F. Ravndal, Phys. Rev. D {\bf  2}, 1278 (1970).
\bibitem{Bajpai}
	R. P. Bajpai, Nucl. Phys. B {\bf  26},  231 (1971).
\bibitem{Arndt}
	R. A. Arndt {\it et al.}, Phys. Rev. D {\bf  43},  2131 (1991).
\bibitem{Batinic}	
	M. Batini\'c {\it et al.}, Phys. Rev. C {\bf  51},  2310 (1995)
	[arXiv:nucl-th/9501011].
\bibitem{Brodsky}
	S. J. Brodsky and G. R. Farrar, Phys. Rev. Lett. {\bf 31},  1153 (1973).
\bibitem{Matveev}
	V.A. Matveev, R.M. Muradian and A.N. Tavkhelidze, Lett. Nuovo Cim.
	{\bf 7}, 719 (1973).	
\bibitem{Byckling}
	E. Byckling and K. Kajantie, Particle Kinematics, {\it ed. John Wiley
	and Sons, 1973}.
\bibitem{Huang1}
	H.W. Huang and P. Kroll, Eur. Phys. J. C {\bf 17}, 423 (2000) 
	[arXiv:hep-ph/0005318]
\bibitem{Huang2}
	     H.W. Huang, R. Jakob, P. Kroll and  K. Passek-Kumericki, Eur. Phys.
	J. C {\bf 33}, 91 (2004) [arXiv:hep-ph/0309071].
\bibitem{Meyer}
	H.O. Meyer and J. Niskanen, Phys. Rev. C {\bf 47},  2474  (1993).
\bibitem{Duncan}
	F. Duncan {\it et al.}, Phys. Rev. Lett. {\bf 80},  4390 (1998).
\bibitem{Hann}
	H. Hann {\it et al.},	Phys. Rev. Lett. {\bf 82}, 2258 (1999).
\bibitem{Calen}
	H. Cal{\'e}n {\it et al.}, Phys. Rev. C {\bf 58},  2667 (1998).
\bibitem{Zlomanczuk}
	J. Zlomanczuk, {\it AIP Conf. Proc.} {\bf 603},  211 (2001).
\bibitem{Moskal}
	P. Moskal, arXiv:hep-ph/0408162.
\bibitem{SibirtsevK}
	A. Sibirtsev, J. Haidenbauer, S. Krewald and Ulf-G. Mei{\ss}ner,
	J. Phys. G {\bf 32}, R395 (2006) [arXiv:nucl-th/0608028].	
\bibitem{TOF}
         M. Abdel-Bary et al., Eur. Phys. J. A {\bf 29}, 353 (2006).
\bibitem{Sternemann}
	H. J. Besch, F. Krautschneider, K. P. Sternemann and  W. Vollrath, 
	Z. Phys. C {\bf 16}, 1 (1982).
\bibitem{Scheffler}
	P. E. Scheffler and  P. L. Walden, Nucl. Phys. B {\bf  75}
	 125 (1974).
\bibitem{Benz}	
	 P. Benz {\it et al.},  Nucl. Phys. B {\bf  65}, 158 (1973).
\bibitem{Boyarski2}
	A. M. Boyarski {\it et al.}, Phys. Rev. Lett. {\bf 21},  1767 (1968).
\bibitem{Baldin}
	A. Baldin, Nuovo Cim. {\bf 8},  569 (1958).
\bibitem{Neugebauer}
	G. Neugebauer, W. Wales and R.L. Walker, Phys. Rev. {\bf 119},
 	1726 (1960).
\bibitem{Wahlig}
	M.A. Wahlig {\it et al.}, Phys. Rev. Lett. {\bf 13},  103 (1964).
\bibitem{Citron}
	A. Citron {\it et al.}, Phys. Rev. Lett. {\bf 13},  205 (1964).
\bibitem{Said2}
	 Z. Li, R. A. Arndt, L. D. Roper and R. L. Workman, Phys. Rev. 
	C {\bf  47}, 2759 (1993) .
\bibitem{Goity1}
	J.L. Goity and N.N. Scoccola, arXiv:hep-ph/0701244.    
\bibitem{Goity2}
	J.L. Goity and N.N. Scoccola, Phys. Rev. D {\bf 72} 034024 (2005)
        [arXiv:hep-ph/0504101].
\bibitem{Afanasev}
	A. Afanasev, C.E. Carlson and C. Wahlquist, Phys. Lett. B {\bf
	398},  393 (1997).
\bibitem{Pee}
	H. van Pee {\it et al.}, Eur. Phys. J. A {\bf 31}, 61 (2007)
	[arXiv:nucl-ex/07041776].
\end{thebibliography}
\end{document}